\documentclass[%
 reprint,
 amsmath,
 amssymb,
 aps,
 amsart
 linenumbers,
 article
]{revtex4-2}
\makeatletter

\makeatletter
\def\@fnsymbol#1{%
  \ifcase#1
  \or $@$         
  \or $@$           
  \or $@$            
  \or $\dagger$         
  \or $\ddagger$        
  \or $\S$              
  \or $\P$              
  \or $\parallel$       
  \or $\ast\ast$        
  \else
    \@ctrerr
  \fi
}

\makeatother
\usepackage{graphicx}
\usepackage[caption=false]{subfig}
\usepackage{dcolumn}
\usepackage{bm}
\usepackage{braket}
\usepackage{booktabs}
\usepackage{physics}
\usepackage{enumitem}
\usepackage{bbold}
\usepackage{multirow}
\usepackage{ragged2e}
\usepackage[table]{xcolor}
\usepackage[T1]{fontenc}
\usepackage{microtype}
\usepackage{orcidlink}
\usepackage{placeins}
\usepackage[mathlines]{lineno}
\usepackage{etoolbox}
\usepackage{hyperref}
\hypersetup{
    hidelinks
}

\AtBeginEnvironment{figure}{\nolinenumbers}
\AtBeginEnvironment{figure*}{\nolinenumbers}

\begin{document}

\title{Qutrits for physics at the LHC}

\author{Miranda Carou Lai\~no\, \orcidlink{0009-0004-9505-2184} $^{1,3,\ast,\dagger}$}
\email{Miranda.carou@ific.uv.es}
\author{Veronika Chobanova\, \orcidlink{0000-0002-1353-6002} $^{2,\dagger}$}
\email{v.chobanova@udc.es}

\author{Miriam Lucio Mart\'inez\, \orcidlink{0001-6823-2607} $^{3,\dagger}$}
\email{milumar@ific.uv.es}
\address{$^1$Instituto de Física Corpuscular (IFIC), University of Valencia and CSIC\\
 Carrer del Catedràtic José Beltrán Martínez, 2. Paterna, Valencia, Spain.}
\address{$^2$University of A Coruña, CITENI, Campus Industrial de Ferrol, Campus de Esteiro, \\ Rúa Doctor Vázquez Cabrera s/n, 15403 Ferrol, A Coruña, Spain}
\address{$^2$University of A Coruña, Campus da Zapateira, 15008 A Coruña, Spain}
\address{$^{\dagger}$ Equal contributors}
\address{$^{\ast}$ Corresponding author}
\address{ \color{blue}{(\textit{Under-review for Nature, Communications Physics})}}

\begin{abstract}
The identification of anomalous events that are not explained by the
Standard Model of particle physics, and the possible discovery of exotic
physical phenomena pose significant theoretical, experimental and
computational challenges. The task will intensify at the High-Luminosity
Large Hadron Collider and next-generation colliders, such as the proposed
Future Circular Collider. Consequently, considerable challenges are
expected concerning data processing, signal reconstruction, and analysis.
This work explores the previously unstudied application of qutrit-based
Quantum Machine Learning models for anomaly detection in high-energy
physics data, with a focus on LHC scenarios. Motivated by the potential of
higher-dimensional quantum systems (qudits) to enhance state space
capacity and expressive power, we benchmark a qutrit-based quantum
autoencoder against a standard qubit baseline to evaluate its feasibility
and resource requirements. Our results show that the qutrit-based model
achieves competitive anomaly detection performance and higher expressive
power, in terms of anomaly discrimination capacity, while requiring fewer
physical units, thereby demonstrating the practical viability of
Majorana-encoded ternary representations for high-energy physics data
compression within current and near-future quantum frameworks.
\end{abstract}

\maketitle

\section*{\label{sec:Introduction}Introduction\protect}

The High-Luminosity Large Hadron Collider and next-generation particle colliders, such as the proposed Future Circular Collider, pose unprecedented challenges in data acquisition and storage to the field of High Energy Physics (HEP)~\cite{CERN_HiLumi_2023}. In practical terms, this implies a substantial increase in the data volume to be processed in real time compared to the current Run~3 conditions of the Large Hadron Collider (LHC). The exabyte-scale datasets to be stored and analysed will push the limits of traditional tools in particle physics~\cite{Nielsen_2020, Henderson:2021qrr, LHCb_2018roe}. 

Recent advances in machine learning, combined with emerging technologies, open an unprecedented opportunity to exploit the full potential of the available datasets~\cite{Albertsson_2018}. For example, machine learning on edge devices~\cite{heterogeneousedge, machine_learning} and neuromorphic computing~\cite{particles8020040} provide promising solutions for low-latency processing. In addition, Quantum Machine Learning (QML) is being extensively investigated for its unique capability to capture and exploit high-dimensional correlations~\cite{QML_HEP, QML_HEP2, qubits_soa1, qubits_soa2, anomaly_variational, variational_2}.

Although the well-known binary paradigm based on qubits offers potential advantages~\cite{qubits_soa1, qubits_soa2, 1p1q} in terms of efficiency and analytical capacity, as practical applications develop, certain limitations are becoming apparent~\cite{QML_HEP, qubits_soa1, qubits_soa4}. These are related to resource efficiency and expressive power, defined in this context as the ability to represent complex, high-dimensional functions within the available Hilbert space. To address these challenges, multi-level quantum systems, or \textit{qudits}, are explored as potential alternatives~\cite{quditquantum, quditMaps, quditsError, evenBreak, Majorana, trappedIon, ququart, algorth_qudits, qudit_fourier}. These qudit systems include larger Hilbert spaces, which offer greater expressive power for encoding data, more compact variational circuits, and more natural representations for multi-level systems~\cite{quditquantum}, such as neutrinos~\cite{neutrino}.

A qudit quantum system consists of $d\in \mathbb{N} $ levels, generalising the concept of a qubit to dimensions $d > 2$. The qudit state is spanned by $d$ orthogonal basis states, $\big( \ket{0}, \ket{1}, \cdots, \ket{d-1} \big)$~\cite{quditFourier}. This scaling offers a larger state space than an equivalent register of $n$ qubits ($2^n$) since a single qudit can encode $\log_2 d$ classical bits of information, resulting in a higher information storage capacity per qudit.
For example, in 3-level quantum systems, or \textit{qutrits}, the information storage capacity, without using any advanced method of encoding, is $\log_2 3 \approx 1.56$ bits of information, compared to the bit per qubit offered by their binary counterpart~\cite{quditquantum}. This higher storage capacity can, in principle, enable QML algorithms to represent more data with fewer physical systems, thereby reducing the depth of circuits, the number of gates needed to create entanglements, and the cumulative decoherence that devices currently suffer from~\cite{preskill}. 

In addition to the higher compression capacity, qudits provide a richer set of operations for encoding quantum features. This stems from the fact that the SU($d$) algebra, which governs the qudit space,  provides additional generators to build higher-dimensional rotations and feature maps in a $(d^2-1)$-dimensional hypersphere~\cite{quditquantum}. In the case of qutrits, the generators are defined by Gell-Mann matrices, enabling the input of data into an 8-dimensional space and allowing subtle patterns to be captured with fewer parameters~\cite{quditMaps}, therefore improving its expressive power.  

Furthermore, qudits offer a significant advantage in information density and logical efficiency, as their higher dimensionality allows for more resilient error-correcting codes with fewer physical units compared to qubits \cite{quditquantum, quditsError}, even enabling the correction of amplitude-damping noise with higher efficiency for $d \geq 3$ \cite{noisy_2}. In addition, qudits can simplify complex algorithms by using higher energy levels as internal ancillas. For instance, $d=5$ systems can decompose the Toffoli gate with linear instead of exponential complexity, and improve Grover’s algorithm performance~\cite{qd_toffoli}. This architectural advantage is supported by theoretical proofs of unconditional quantum-classical separations, where qudit-based search problems outperform classical polynomial-size circuits~\cite{unconditional}. However, these benefits face a steep trade-off: noise affects qudits more rapidly as the dimension $d$ increases, requiring significantly faster gate operations to remain competitive and leading to a quicker transition into complex, non-linear noise regimes~\cite{Noisy, fidelity_qd}. 

While qudits excel in local fidelity and circuit depth, they currently lag behind qubits in global scalability. Experimental implementations for $d > 3$ remain physically challenging due to restricted level transitions that necessitate extra gates to move information, and the fact that decomposing qudit gates into qubit-based architectures on real hardware inevitably leads to a loss of the depth efficiency that motivates their use~\cite{qd_toffoli, unconditional}. This is why the development of native qudit-based hardware is essential, as it would preserve those advantages and avoid the efficiency loss introduced by such decompositions. A fundamental step in realising such native architectures is to accurately map the qudit's state space.

As established in Refs.~\cite{geometry, KIMURA2003339}, the Bloch sphere used to describe the state space of qubits $(d =2)$ can be generalised to a hypersphere to describe a $d$-level system; nevertheless, this representation does not provide a trivial distinction between pure and mixed states, therefore hindering the technical implementation of such hyperspheres.
This can be achieved within representations of qudits on unit spheres. 
In the example of the Majorana stellar representation~\cite{Majorana1932}, points on the surface are identified as pure states, while the extension of this representation to mixed states associates them with configurations that effectively explore the interior of the unit sphere~\cite{mixed_states}. This geometric distinction is fundamental for a complete description of the qutrit Hilbert space. The Majorana representation thus provides a robust mathematical framework for encoding qutrits, facilitating the mapping of physical quantum states. This feature is less straightforward in other hypersphere generalisations~\cite{geometry, qRepresentation1, qRepresentation2, qRepresentation3}, and motivates our choice of the Majorana representation in this work.

We propose a qutrit-based quantum anomaly detection framework and evaluate its performance against a standard qubit-based baseline. This comparison focuses on two key metrics: detection accuracy and expressive power. Anomaly detection in this context refers to outlier detection, where a Quantum Autoencoder (QAE)~\cite{variational_2, 1p1q} is trained on jet data from the CMS detector~\cite{aspenJets, jetClass} in order to learn the latent representation of standard Quantum Chromodynamics (QCD) background jets. Signal events are identified as anomalous when their reconstruction fidelity falls significantly below the learned norm, marking them as physical outliers relative to the background distribution~\cite{ml_ad_hep}. To adapt the QAE architecture to ternary logic, the HEP data is mapped onto a qutrit state space through a Majorana-based encoding. This encoding scheme is mathematically derived from the quantum state formalism of spin-1 systems, and applied here within a classical data setting to map classical jet observables and constituent kinematics into the qutrit Hilbert space. Furthermore, we implement a set of logic gates that generalise conventional two-qubit operations to accommodate the expanded three-level Hilbert space.

While the Majorana geometric representation for qutrits has been previously established in the context of spin-1 systems and NMR implementations~\cite{Majorana}, its application to high-energy physics remains largely unexplored. In this work, we address this gap by developing an algorithm for LHC anomaly detection that employs Majorana-encoded qutrits.  By adopting the Majorana geometric representation, originally used for physical spin-1 systems, as a mathematical encoding scheme, we construct a practical framework to map jet kinematics. This allows us to leverage the expanded $3^N$ state space to represent multidimensional features
 and systematically assess the practical advantages of ternary architectures for large-scale collider data, providing a specialised infrastructure that links theoretical qutrit dynamics with the specific computational requirements of future collider experiments. Specifically, by fixing the number of quantum systems $N$, the expanded $3^N$ Hilbert space is utilised to encode jet kinematic parameters, thereby enabling an exploration of how this increased representational capacity influences anomaly detection performance. In the present context, the qutrit representation functions as an enhanced computational framework for classical feature mapping.
\section*{\label{sec:Methods}Methods\protect}
\subsection*{\label{sec:reference}Baseline work used for benchmarking}
In order to implement a robust qutrit-based model for anomaly detection, we adopt an established quantum autoencoder as our baseline model for benchmarking. This model introduces a quantum encoding scheme for particle physics, the “One Particle–One Qubit” (1P1Q) encoding~\cite{1p1q}, in which each particle’s kinematics are encoded into a single qubit without prior classical compression. To accomplish this, the pseudorapidity $\eta$ and azimuthal angle $\phi$, modulated by the transverse momentum, $p_T$, and normalised by the jet transverse momentum, $p_{T,\text{jet}}$, are mapped onto the spherical coordinates of the Bloch sphere (see Ref.~\cite{1p1q} for details).

To validate their novel encoding scheme, the authors applied it to two different models: a QAE for unsupervised anomaly detection and a Variational Quantum Circuit (VQC) for supervised classification of top-quark jets. In both cases, they performed a systematic benchmarking against the corresponding classical versions of the models. Since their QAE achieves competitive performance compared to the classical autoencoder (CAE) across all signal types, we set this quantum model as our baseline for comparison and use their encoding scheme for particle physics as a starting point in the design of our qutrit-based encoding.

\subsection*{\label{sec:QAE} Quantum autoencoder for anomaly detection}
The QAE model under consideration and used in Ref.~\cite{ 1p1q} is a well-known algorithm, widely used by the community for the task of anomaly detection~\cite{qae, variational_2, Frehner2025, qae_security, Sakhnenko2022} and involves the implementation of two structures, an encoder and a decoder, through variational quantum circuits. The encoder -- also referred to as \textit{variational layer} -- compresses the input data into a reduced latent representation, discarding redundant qubits (or qutrits). The discarded qubits (qutrits) are referred to as {\it trash states}. The decoder then reconstructs the input from this latent state using the Hermitian conjugate of the encoding operators. 
\begin{figure}[h]
    \centering
    \includegraphics[width=8cm]{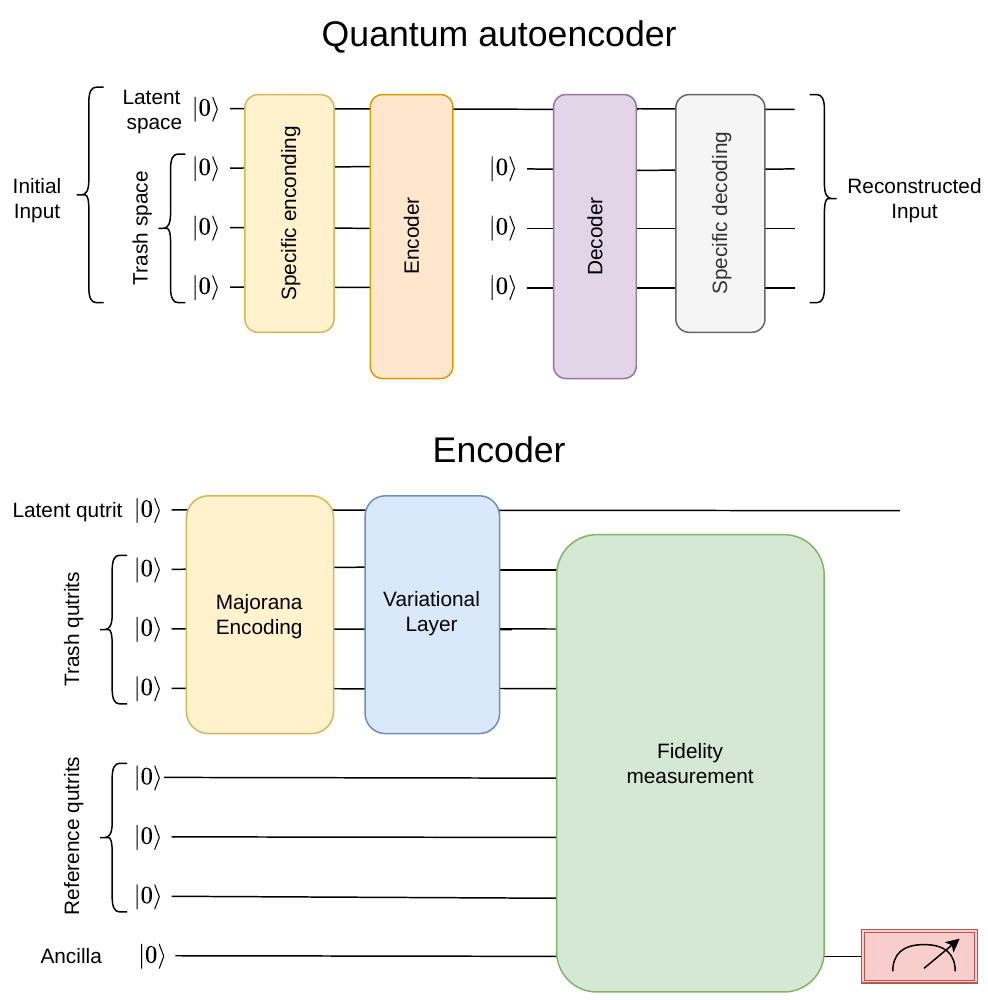}
    \caption{\footnotesize \justifying \textbf{General quantum autoencoder and encoder structure.} The top panel illustrates the full QAE scheme, consisting of encoding, compression into a latent space, and decoding phases. The bottom panel details the encoder-only architecture used during training, where the trash space state is compared against a reference state to perform the fidelity measurement via an ancilla.} 
    \label{fig:qae_general}
\end{figure}

In this work, only the fidelity from the encoder $U(\Theta)$, is used as an anomaly indicator. The encoder is defined as
\begin{eqnarray}
U(\Theta) &=&
\left( \bigotimes_{i=1}^{N}
R_X(\varsigma_i) R_Y(\upsilon_i) R_Z(\gamma_i)
\right) \nonumber\\
&& \otimes
\left( \bigotimes_{1 \le i < j \le N} C_{ij} \right)
\label{eq:one}
\end{eqnarray}
where $\varsigma_i$, $\upsilon_i$, and $\gamma_i$ are trainable parameters, $C_{ij}$ denotes the two-qubit CNOT gates, and $R_X$, $R_Y$, and $R_Z$ are the rotation gates around the $X$, $Y$, and $Z$ axes in a 2-level quantum system (see Ref. \cite{nielsen_chuang_2010} for more details). To update the trainable parameters, the Adam optimizer~\cite{adam}, a stochastic gradient descent method based on the adaptive estimation of first-order and second-order moments, is employed. A learning rate of $0.05$ is set, and the AMSGrad variant~\cite{amsgrad} is utilised to improve convergence properties by maintaining the maximum of past squared gradients. As illustrated in Fig.~\ref{fig:qae_general}, the operator $U(\Theta)$ acts on both the latent and trash states, performing the compression of the input state.

The cost function is defined as the negative fidelity between the \textit{reference} states, initialised as $\ket{0}$, and the \textit{trash} states (see Refs .~\cite{qae, 1p1q} for details). To perform the fidelity measurement, a SWAP test is used to quantify how effectively the input information has been compressed into the latent space. The SWAP test was first introduced in Ref.~\cite{finger_quantum} and is a fundamental quantum algorithm used to measure the similarity (specifically, the overlap) between two quantum states, without requiring explicit knowledge of the states themselves (see Ref.~\cite{finger_quantum} for more details). In this implementation, it serves as an anomaly indicator, as it quantifies how effectively the model compresses information into the latent space through the variational layer. 
\subsection*{\label{sec:dataset} Dataset}
The JetClass~\cite{jetClass} dataset and the data collected in 2016 by the CMS detector at the LHC and made public~\cite{aspenJets} have been used separately for the optimal training of the model in different scenarios. The 2016 CMS dataset contains approximately 178 million  high-$p_T$ jets stored in a format suitable for Machine Learning. The data is dominated by Quantum Chromodynamics (QCD) jets with less than 1\% contamination from other sources. The JetClass contains 125 million jets, divided into ten classes, as outlined in Ref.~\cite{jetClass}, which are split across training, validation, testing, and inference.
 The specific classes utilised in this study are defined as follows:
\begin{enumerate}[noitemsep, topsep=1pt, partopsep=1pt, parsep=1pt]
    \item $H \to b\bar{b}$: Higgs boson decays into a beauty quark–antiquark pair.
    \item $t \to b q \bar{q}$: Top quark decays into a beauty quark and two light quarks.
    \item $W \to q\bar{q}$: $W$ boson decays into two quarks.
    \item $q/g$: Jets initiated by a single quark or gluon, representing generic QCD background processes.
\end{enumerate}

\begin{figure}[h]
\includegraphics[width=8cm]{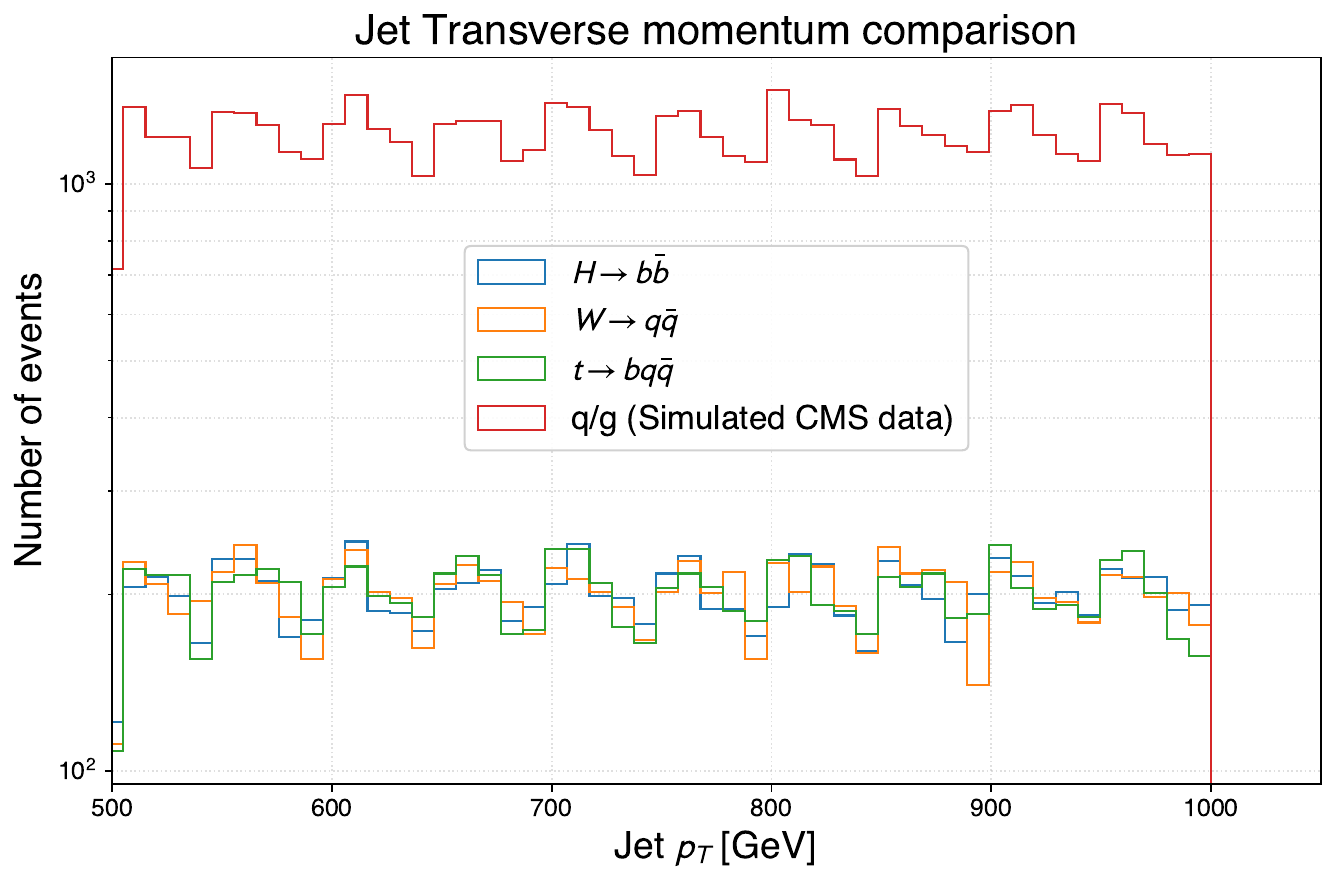}
\caption{\protect\footnotesize \protect\justifying \textbf{Histogram showing the flat $p_{T,\text{jet}}$ distribution of the different signals between $500$ and $1000$~GeV in the simulated CMS dataset.} A larger number of background events than signal events is included, as they are used during the training and testing phases.}
\label{fig:JETPT}
\end{figure}

During the inference phase, when the trained model applies its learned patterns to compress previously unseen data, the simulated dataset JetClass~\cite{jetClass} is always employed. For the present analysis, we restrict ourselves to the signals $H \to b\bar{b}$, $t \to b q q'$, $W \to q q'$, and $q/g$, matching the categories used in the baseline model~\cite{1p1q}. This choice ensures a consistent comparison framework and allows for a more reliable assessment of the performance of our qutrit-based architecture in distinguishing signal from background jets.

The training phase of the QAE model is performed in some cases with experimental data and in others with simulated data. To construct the training sample, events are selected such that each class follows the same jet transverse momentum distribution, flat in $p_{T,\text{jet}}$ in the range $[500,1000] \textrm{ GeV}$, as shown in Fig.~\ref{fig:JETPT}, so that each class draws from a uniform distribution. This choice prevents the model from exploiting differences in the overall jet energy scale, characterised by $p_{T,\text{jet}}$, and instead forces it to focus exclusively on jet substructure, following the strategy of Ref.~\cite{1p1q}. In this work, jet substructure refers to the internal distribution of constituents and their correlations within the jet cone, which provide the discriminating features used to identify the jet origin.
\subsection*{Metrics used for the evaluation of the model performance}
In the present study, the $\text{AUC}$ is consistently derived from signal-background discrimination. Specifically, the background samples are drawn from the same source used during training: JetClass data when the model is trained on simulated data and CMS 2016 open data otherwise. The anomaly score is defined as $1 - \mathcal{F}$, where $\mathcal{F}$ represents the fidelity obtained via the SWAP test. Performance is then quantified by evaluating the $\text{AUC}$ from the resulting Receiver Operating Characteristic (ROC) curves~\cite{FAWCETT2006861}  in accordance with standard classification benchmarks. 

This is not the only metric employed to assess the model: the Jensen–Shannon distance and the Wasserstein distance (see Sections D and E in the Supplementary Information document) are also used to quantify its expressive power. The latter distance accounts for the underlying metric geometry by measuring the cost of transforming one distribution into another, and it is particularly sensitive to global shifts and shape deformations in the fidelity distributions. Since the autoencoder is trained exclusively on background events, any deviation in the physical jet substructure (such as variations in energy distributions or multi-prong decay features) leads to systematic reductions in compression fidelity. Consequently, a larger Wasserstein distance between fidelity distributions quantitatively reflects the model's capacity to isolate subtle physical modifications in the jet substructure within the latent space.

To ensure the statistical robustness of our results, all reported metrics are derived from a set of 100 independent runs. In this context, an `independent run' refers to a full execution of the analysis pipeline, including the training, validation, and testing phases. For each run, a consistent number of events is assigned to each phase: 10,000 events for training, 2,500 for validation, and 10,000 for each signal and background sample in the inference phase. Prior to partitioning, the complete dataset is shuffled, and events are randomly sampled, with a unique random seed assigned to each run. This procedure guarantees the statistical independence of each run and mitigates potential biases in model evaluation.

For each iteration, the Wasserstein and Jensen–Shannon distances are computed, as well as the ROC curves and the AUC corresponding to each signal. The final reported values correspond to the mean ($\mu$), and we take the standard deviation ($\sigma$) over these 100 samples as the statistical uncertainty. This not only allows us to assess the nominal ability of the model to discriminate signals ($H\to b\bar{b}$, $W \to q\bar{q}$, $t \to bq\bar{q}$) from the QCD background, but also to quantify the stability and reproducibility of the quantum autoencoder in the presence of fluctuations in the reconstruction.

 To ensure precision, the quantum circuit's performance is evaluated via ideal state vector simulations rather than stochastic sampling. In practical terms, this corresponds to an exact analytical evaluation, which eliminates statistical fluctuations or shot noise, and each execution yields the exact probability amplitudes of the ancilla qutrit. From these amplitudes, the fidelity is retrieved directly as the probability of the zero-state, $\mathcal{F} = P(\ket{0})$.
\subsection*{\label{sec:Majorana}Majorana stellar representation for qutrits}
The Majorana formalism introduced in the Introduction section serves as the foundation for our encoding and for the majority of the generalised gates employed in our model.
\begin{figure}[h!]
\includegraphics[width=8.7cm]{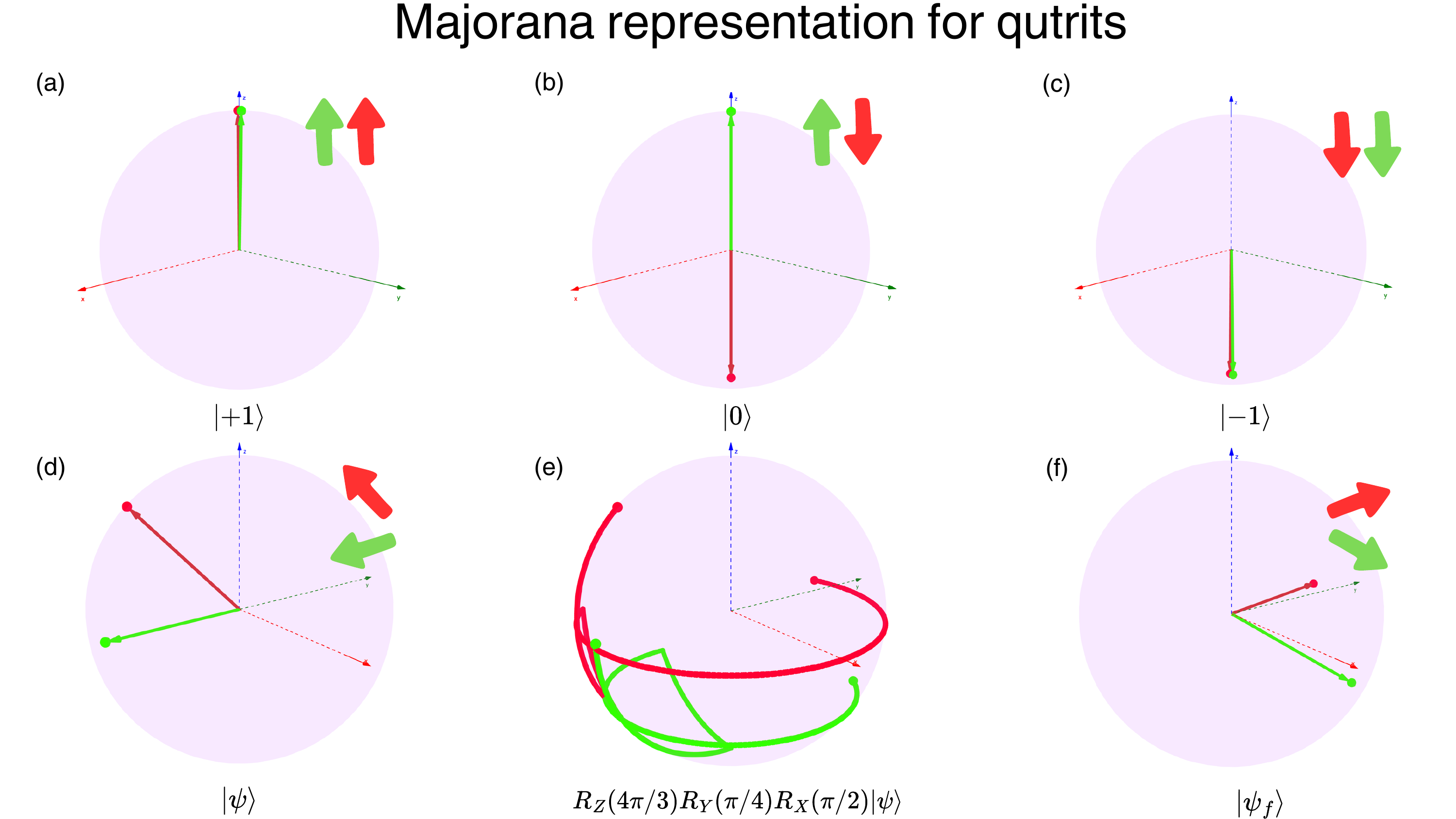}
\caption{\protect\footnotesize \protect\justifying \textbf{Majorana representation of a qutrit.} The state is defined by two points, or "Majorana stars", $P_1$ (green point) and $P_2$ (red point) on the unit sphere's surface. The basis eigenstates are represented by specific configurations: (a) the $\ket{+1}$ state corresponds to both points coinciding at the North Pole; (b) the $\ket{0}$ state is defined by one point at the North Pole and the other at the South Pole; and (c) the $\ket{-1}$ state is represented by both points coinciding at the South Pole. (d) A qutrit pure state, $\ket{\psi}$, represented by two points and encoded using a set of four angles defined by $\theta_1 = \frac{\pi}{2}$, $\theta_2 = \frac{\pi}{4}$, $\phi_1 = \frac{3\pi}{2}$, and $\phi_2 = \frac{4\pi}{3}$. (e) The trajectories of both points representing $\ket{\psi}$ under the application of the rotation gates $R_X(\theta_1)$, $R_Y(\theta_2)$, and $R_Z(\phi_2)$. (f) The final state, $\ket{\psi_f}$, obtained after applying the rotation gates to $\ket{\psi}$.}
\label{fig:epsart}
\end{figure}
\subsubsection*{ \label{sub:gell-man} Gell-Mann matrices}
The qutrit generators are the eight Gell-Mann matrices, denoted by $\lambda_i$ ($i\in \{1,\cdots,8\}$), 
\begin{eqnarray}
 \nonumber &\lambda_1 = \begin{bmatrix}
0 & 1 & 0 \\
1 & 0 & 0 \\
0 & 0 & 0
\end{bmatrix} \,\,\,
\lambda_2 = \begin{bmatrix}
0 & -\mathrm{i} & 0 \\
\mathrm{i} & 0 & 0 \\
0 & 0 & 0
\end{bmatrix} \,\,\,
\lambda_3 = \begin{bmatrix}
1 & 0 & 0 \\
0 & -1 & 0 \\
0 & 0 & 0
\end{bmatrix} \\ \nonumber \\\nonumber 
 &\lambda_4 = \begin{bmatrix}
0 & 0 & 1 \\
0 & 0 & 0 \\
1 & 0 & 0
\end{bmatrix}  \,\,\,
\lambda_5 = \begin{bmatrix}
0 & 0 & -\mathrm{i} \\
0 & 0 & 0 \\
\mathrm{i} & 0 & 0
\end{bmatrix} \quad
\lambda_6 = \begin{bmatrix}
0 & 0 & 0 \\
0 & 0 & 1 \\
0 & 1 & 0
\end{bmatrix} \\ \nonumber \\ 
&\lambda_7 = \begin{bmatrix}
0 & 0 & 0 \\
0 & 0 & -\mathrm{i} \\
0 & \mathrm{i} & 0
\end{bmatrix} \quad\quad
\lambda_8 = \frac{1}{\sqrt{3}} \begin{bmatrix}
1 & 0 & 0 \\
0 & 1 & 0 \\
0 & 0 & -2
\end{bmatrix},
\label{eq:gellman}
\end{eqnarray}
which play a role analogous to the Pauli matrices for qubits. The Gell-Mann matrices form a complete Hermitian set of generators for the unimodular group SU(3), which is the symmetry group of a three-level quantum system, and play a fundamental role in the study of qutrits~\cite{geometry}. The transformations generated by the complex exponentiation of the Gell-Mann matrices are given by 
\begin{equation}
U_{\lambda_i}(\theta) = e^{i\theta \lambda_i}.
\label{eq:expoGell}
\end{equation}
The $U_{\lambda_i}(\theta)$ matrices represent different types of unitary transformations in a specific subspace within the entire qutrit state, except for $U_{\lambda_8}(\theta)$, which only adds a relative phase. The set of SO(3) generators, $\Sigma_i$ ($i\in \{1,2,3\})$, are defined as
\begin{eqnarray}
    \Sigma_1 &= \frac{1}{\sqrt{2}}
    \begin{bmatrix} 
    0 & 1 & 0 \\ 
    1 & 0 & 1 \\ 
    0 & 1 & 0 \end{bmatrix} \quad
    \Sigma_2 = \frac{1}{\sqrt{2}} 
    \begin{bmatrix} 
    0 & -i & 0 \\ 
    i & 0 & -i \\ 
    0 & i & 0 
    \end{bmatrix} \nonumber\\\nonumber\\ \nonumber
    &\Sigma_3 = \begin{bmatrix} 
    1 & 0 & 0 \\ 
    0 & 0 & 0 \\ 
    0 & 0 & -1 
    \end{bmatrix}.  \\
    \label{eq:transform}
\end{eqnarray}
The $\Sigma_i$ matrices generate the corresponding three-dimensional physical rotation around the $j$-th Cartesian axis ($j \in \{X,Y,Z\}$). It should be noted that the matrices of the $\Sigma_i$  group are a normalised linear combination of two Gell-Mann matrices. For instance, $\Sigma_1$  is the result of the sum of $\lambda_1$ plus $\lambda_6$ multiplied by a normalisation factor $\frac{1}{\sqrt{2}}$. In this way, the operator $\Sigma_1$  can act as a unitary transformation within the space $\{\ket{-1},\ket{0},\ket{1}\}$. The points on the Majorana sphere transform as vectors in three-dimensional space under rotations~\cite{Majorana}. Thus, the $\Sigma_i$  group produces real physical rotations, moving the two points of the Majorana sphere in $\mathbb{R}^3$. 
\subsubsection*{Qutrits representation}
The representation of spin-$s$ systems is defined by $2s + 1$ orthonormal eigenvectors of the angular momentum operator $L_z$. These eigenvectors $\{\ket{j}\}$ with $j$ between $-s$ and $+s$ provide the basis for the Hilbert space of spin-$s$ states.
In the specific case of qutrits ($s=1$), the three-dimensional Hilbert space is described by three eigenvectors $\{\ket{-1},\ket{0},\ket{1}\}$ of the angular momentum operators $\sum_3(L_z)$. Therefore, the general state is determined by
\begin{equation}
    \ket{\psi} = C_{-1} \ket{-1} + C_0\ket{0}+ C_{+1}\ket{+1}.
    \label{generalstate}
\end{equation}
By examining the representation of the pure states within the Majorana's sphere formalism~\cite{Majorana, Majorana1932}, it becomes evident that a qutrit state can be fully characterised by two points within the unit sphere, $P_1(\theta_1, \phi_1)$ and $P_2(\theta_2, \phi_2)$, where $\theta_i \in [0,\pi]$ and $\phi_i \in [0,2\pi]$. The Majorana representation of the different eigenstates of a pure qutrit state, as well as the behaviour of a specific qutrit state under rotation gates, are shown in Fig.~\ref{fig:epsart}.

A general state, $\ket{\psi_g}$, can be explicitly written in terms of the four angles $\theta_1, \theta_2, \phi_1$ and $\phi_2$, as
\begin{subequations}
\begin{eqnarray}
 \ket{\psi_g}= \Gamma
\begin{bmatrix}
\sqrt{2} \cos \dfrac{\theta_1}{2} \cos \dfrac{\theta_2}{2} \\
e^{i\phi_1} \sin \dfrac{\theta_1}{2} \cos \dfrac{\theta_2}{2} + e^{i\phi_2} \cos \dfrac{\theta_1}{2} \sin \dfrac{\theta_2}{2} \\
\sqrt{2} e^{i(\phi_1 + \phi_2)} \sin \dfrac{\theta_1}{2} \sin \dfrac{\theta_2}{2}
\end{bmatrix} 
\end{eqnarray}
\text{where} 
\begin{eqnarray}
\Gamma &=& \sqrt{2}
\left[
3 + \cos \theta_1 \cos \theta_2
\right. \nonumber\\
&& \left.
+ \sin \theta_1 \sin \theta_2
\cos(\phi_1 - \phi_2)
\right]^{-\frac{1}{2}}
\end{eqnarray}
\label{eq:encodingM}
\end{subequations}
is a normalisation factor (see Section B of the Supplementary Information document for details on the derivation). Note that a general pure state in a three-level system is initially defined by three complex amplitudes, $C_i$, which correspond to six real parameters. However, the physical state is subject to a fundamental constraint: the normalisation condition $\sum_{i=1}^3 |C_i|^2 = 1$. Moreover, one of the parameters is an overall global phase. This reduces the effective degrees of freedom to four independent real angles, which uniquely parametrise the space of pure qutrit states. More details can be found in Ref.~\cite{Majorana} and Ref.~\cite{Majorana1932}. It is important to emphasise here that the Majorana representation serves as an encoding framework applied to classical data. The input parameters mapped onto the four angles $(\theta_1, \theta_2, \phi_1, \phi_2)$ in Eq.~\eqref{eq:encodingM} do not correspond to pre-existing physical quantum states. Rather, they are classical observables embedded within the coordinates of the Majorana stellar representation. In this context, the quantum state structure is employed as a mathematical construct to capture multi-dimensional classical correlations.
\subsection*{ Generalized qutrit gates} 
Building on the generalisation of the rotation gates to the qutrit space previously explained in the Majorana Stellar Representation for Qutrits subsection of the Methods section, we present the generalisation of two other essential gates required for the implementation of a qutrit-based QAE model: the Hadamard and SWAP gates.
\subsubsection*{ Hadamard gate }
The Chrestenson gate (Ch) presented in Ref.~\cite{Majorana}, and the ternary Hadamard (THadamard) gate~\cite{thadamard1, thadamard2} implemented in the QML library  PennyLane~\cite{PennyLane} and used in the development of our algorithm are equivalent. Both serve as a qutrit analogue to the Hadamard gate for qubits. Like its binary counterpart, the Ch gate maps each computational basis state into an equal superposition of all three qutrit levels, with relative phases. Its matrix representation is given by
\begin{equation}
    Ch = \frac{1}{\sqrt{3}} 
\begin{bmatrix} 
1 & 1 & 1 \\ 
1 & e^{2\pi i/3} & e^{4\pi i/3} \\ 
1 & e^{4\pi i/3} & e^{2\pi i/3} 
\end{bmatrix}    .
\label{eq:ch}
\end{equation}
\subsubsection*{\label{sec:SWAP}SWAP gate }
In the present study, the generalised SWAP gate for qutrits (see Ref.~\cite{SWAP} for details) is utilised. It is given by
\begin{equation}
    U_{\text{SWAP}} =
\begin{bmatrix}
1 & 0 & 0 & 0 & 0 & 0 & 0 & 0 & 0 \\
0 & 0 & 0 & 1 & 0 & 0 & 0 & 0 & 0 \\
0 & 0 & 0 & 0 & 0 & 0 & 1 & 0 & 0 \\
0 & 1 & 0 & 0 & 0 & 0 & 0 & 0 & 0 \\
0 & 0 & 0 & 0 & 1 & 0 & 0 & 0 & 0 \\
0 & 0 & 0 & 0 & 0 & 0 & 0 & 1 & 0 \\
0 & 0 & 1 & 0 & 0 & 0 & 0 & 0 & 0 \\
0 & 0 & 0 & 0 & 0 & 1 & 0 & 0 & 0 \\
0 & 0 & 0 & 0 & 0 & 0 & 0 & 0 & 1
\end{bmatrix}.
\label{eq:swap} 
\end{equation}
This gate acts as a permutation gate, capable of interchanging the states of a pair of qutrits.
The structure under consideration comprises three fixed elements and a set of three transpositions.
\subsection*{\label{sec:ProposedModel}Implementation with qutrits}
\subsubsection*{Model adjustments and logic gates}
When extending the core idea behind the 1P1Q encoding scheme to qutrits, that is, assigning one particle per qubit or, in our case, one particle or jet per qutrit, the general structure of the QAE remains largely unchanged, as it follows the well-established framework of the algorithm from Ref.~\cite{1p1q}. Nevertheless,  significant modifications to the rotation gates and encoding schemes are necessary to align with the qutrit paradigm. 

\begin{figure}[h]
\includegraphics[width=8cm]{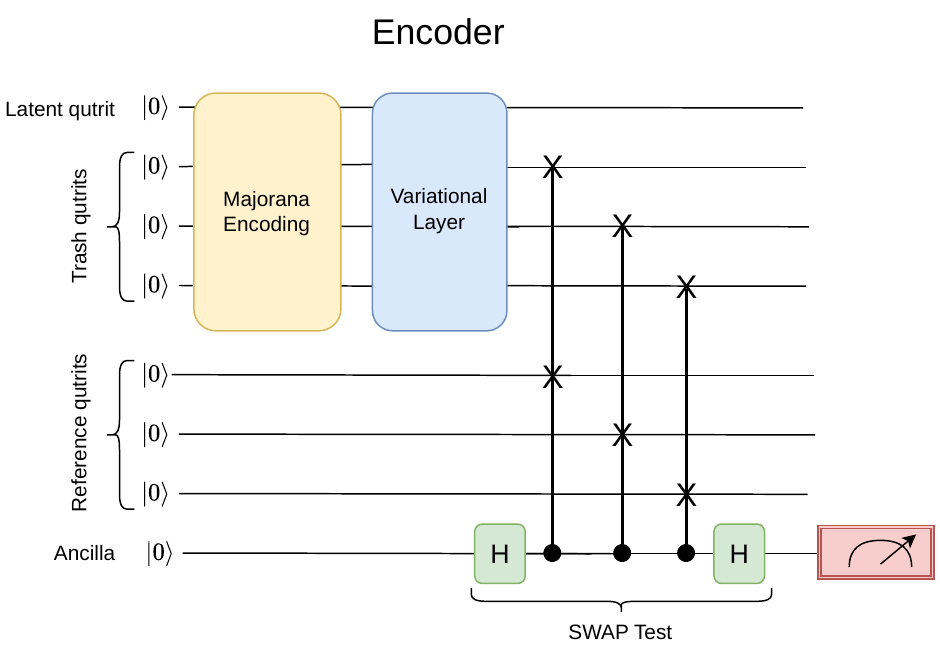}
\caption{\footnotesize \justifying
\textbf{Qutrit-based QAE circuit diagram.} QAE model with a latent space of one qutrit, along with three \textit{trash} qutrits, three \textit{reference} qutrits, and a single ancilla qutrit. The structure represents the implementation of the Majorana encoding, followed by the adapted variational layer and ending with the implementation of the SWAP test.}
\label{fig:circuit}
\end{figure}

First, the layout of most of the operations and logic gates is replaced with operators suited to the new requirements. The declaration of new rotation gates following Eq.~(\ref{eq:transform}) based on the Gell-Mann matrices defined in Eq.~(\ref{eq:gellman}), is an essential part of this process. The application of rotations around the the $j$-th Cartesian axis ($j \in \{X,Y,Z\}$) using the $\Sigma_i$ matrices ($i\in \{1,2,3\}$)  by an angle $\xi$ within the sphere, in conjunction with the movement of both points of the Majorana representation~\cite{Majorana}, is achieved using the exponentiation formula according to
\begin{eqnarray}
    R_i(\xi) &= e^{\, i \xi \Sigma_i}
= I + (\cos \xi - 1)\,\Sigma_i^2 + i \sin \xi \,\Sigma_i
\label{eq:rotation}.
\end{eqnarray}
The qutrit Controlled-SWAP (CSWAP) gate is employed in this work to implement the SWAP test. It is realized using PennyLane’s  \texttt{`ControlledQutritUnitary'} operation, ensuring that the state of the target qutrit is exchanged when the control qutrit occupies the $\ket{2}$ state (corresponding to $\ket{1}$ in spin notation), whereby a single control qutrit triggers a generalized ($U_{\text{SWAP}}$) operation. Other gates, such as the generalisation of the CNOT gate and the Chrestenson gate (see Eq.~(\ref{eq:ch})), do not require manual implementation since they are available in PennyLane. The general structure of the model is illustrated in Fig.~\ref{fig:circuit}.
 
As a second step, the initial 1P1Q encoding scheme is replaced by a scheme based on Majorana encoding, following Eq.(\ref{eq:encodingM}). While we retain the essence of the 1P1Q strategy by mapping one particle or jet per qutrit, we exploit the additional degrees of freedom of the qutrit Hilbert space to go beyond the basic four-momentum components used for the particle encoding in the qubit baseline. Specifically, the Majorana representation allows us to incorporate two extra angular parameters that encode further jet-structure information, resulting in a qutrit state defined by four independent angles. Further details regarding the implementation of the Majorana encoding can be found in sections B and C of the Supplementary Information document.

To evaluate the interdependence within the JetClass dataset, a correlation analysis was performed between the global jet features and the individual particle-level constituents. Based on the observed degrees of correlation, the $N$-subjettiness ratios, jet energy, and the longitudinal and transverse impact parameters were identified as parameters of interest. The $N$-subjettiness observable, originally proposed in Ref.~\cite{taus}, is defined as
\begin{subequations}
\begin{eqnarray}
\tau_N = \frac{1}{d_0} \sum_k p_{T,k}\, \min \{ \Delta R_{1,k},\, \Delta R_{2,k},\, \ldots,\, \Delta R_{N,k} \},\nonumber\\
\end{eqnarray}
\end{subequations}
where the sum runs over the jet constituents and $\Delta R_{J,k}$ denotes the distance in the rapidity--azimuth plane between the subjet axis $J$ and particle $k$. Here, a subjet is understood as a localized cluster of particles within a jet, corresponding to one of its hard prongs. The subjet axes $J$ therefore represent the directions of these prongs. The N-subjettiness quantifies how consistent a jet is with having $N$ subjets. It is particularly useful in the study of boosted hadronic objects, as it captures their multi-prong structure. Ratios such as $\tau_2/\tau_1$ and $\tau_3/\tau_2$ are widely used to discriminate between two-prong (e.g. W, Z, Higgs) and three-prong (e.g. top quark) topologies. Further details can be found in section F of the Supplementary Information document and in Ref.~\cite{taus}.

The 1P1Q encoding works such that it can be extended to include essential jet-level information and features that refer to the jet as a whole, such as the N-subjettiness. In this sense, the resulting construction is no longer a 1P1Q encoding in the strict sense, but rather a one-particle-one-qutrit with jet context scheme. Expanding upon the original encoding (1P1Q), we included parameters describing the main features from each jet, such as the jet energy ($E$), the ratio between pairs of $\tau_N$ where $N\in(1,..,4)$, defined as
\begin{subequations}
\begin{eqnarray}
\tau_{ij}' = \frac{\tau_j}{\tau_i} \quad\quad i,j \in (1,..,4),
\end{eqnarray} 
\label{eq:taus}
\end{subequations}the longitudinal impact parameter ($d_0$), and the transverse impact parameter ($d_z$). They are rescaled according to
\begin{subequations}
\begin{eqnarray}
f&=& {\big(}1 + \frac{2\pi}{1 + e^{-w}}{\big)}\cdot\frac{p_T}{P_{T,\text{jet}}} \label{eq:f}\\\nonumber\\ 
\varphi &=&f \cdot (\phi - \phi_{\text{jet}})  
\\
\vartheta &=& f  \cdot (\eta - \eta_{\text{jet}}) 
\\
\varepsilon &=& f \cdot (E - E_{\text{jet}}) 
\label{eq:enco_b}\\
\tau_{12} &=& f \cdot \tau_{12}' 
\label{eq:enco_c}\\
\tau_{23} &=& f \cdot \tau_{23}'
\label{eq:enco_d}\\
\tau_{34} &=& f \cdot \tau_{34}'
\label{eq:enco_e}\\
\varrho_0 &=& f  \cdot d_0
\label{eq:enco_f}\\
\varrho_z &=& f \cdot d_z,
\label{eq:enco_g}\\ \nonumber
\end{eqnarray}
\label{eq:newenco}
\end{subequations}
where $\phi_{\text{jet}}$, $\eta_{\text{jet}}$ and $E_{\text{jet}}$ denote the jet’s azimuthal angle, pseudorapidity and energy, respectively.  The factor $f$ is introduced as a scaling parameter with a trainable parameter $\omega$, to ensure that, after the encoding, the particles are not clustered around one of the poles, as discussed in Ref.~\cite{1p1q}. When the subscript “jet” is omitted, the corresponding quantities refer to individual particles that form the jet. 

After the encoding analysis performed (as detailed in Supplementary Information A), the combination that yielded the highest value of Area Under the Curve (AUC) and is hereafter considered as the baseline for qutrits is the one in which $\theta_1 = \vartheta$ and $\theta_2 = \varphi$, as used in Ref.~\cite{1p1q}, remain fixed. As for $\phi_1$ and $\phi_2$, the combinations $\phi_1 \in \{\tau_{12}, \tau_{23}, \tau_{34}\}$, $\phi_2 = \varepsilon$ and $\phi_1 = \varrho_0$, $\phi_2 = \varrho_z$ prove to yield the highest performance. Physically redundant pairings, such as energy $E$ with invariant mass $m$, were systematically excluded since they did not significantly improve performance. The mass is strongly correlated with the four-momentum, and therefore does not provide additional discriminating power in this context. 

It should be noted that when working with simulated CMS events, the three-momentum $(p)$ alone is sufficient to fully describe the kinematic state of each particle. However, when training the QAE with data, truth-level information, such as particle type or simulation labels, is not available. Therefore, energy $(E)$ and $(p)$ must be included as independent inputs to capture additional kinematic information without relying on these labels. This ensures that our parametrisation is not redundant but rather necessary for a more complete description of the particle kinematics. Moreover, longitudinal and transverse impact parameters $(d_0, d_z)$ provide spatial information about the production vertex and the particle trajectory, introducing an additional degree of geometric discrimination in both simulation and real events that is not encoded in the four-momentum~\cite{PARTICLE}.

In its current state, the simulation of quantum circuits with more than eight qutrits in PennyLane requires a substantial amount of memory, making it unfeasible to run a circuit of 19 qutrits on a CPU with less than 300~GB of RAM. As a result, our circuit design has to be reduced to a single-qutrit latent space, three trash qutrits, three reference qutrits, and one ancilla qutrit. Thus, when using eight qutrits, a maximum of four particles can be encoded per execution. To be able to compare them, both the qutrit and the qubit QAE models are scaled down. Moreover, we comment on the asymmetric choice of particle scaling between the qutrit-based quantum model (4-particle encoding) and the classical baseline (10-particle encoding). In Ref~\cite{1p1q}, the classical autoencoder (CAE) is optimised for input vectors of size 30 (10 particles × 3 variables: $p_T, \eta, \phi$) and employs a deep architecture with dense layers of 20--16--12 nodes. Adapting this CAE to a 4-particle encoding (input size $12$) would necessitate complete re-engineering and hyperparameter tuning to avoid severe overfitting and architectural inconsistencies. Consequently, we keep the fully-optimised classical baseline from the literature~\cite{1p1q} as a demanding, asymptotic upper-bound target. This ensures that the competitive performance trends shown by our resource-constrained qutrit Majorana QAE are evaluated against a highly capable classical counterpart, rather than an artificially weakened and trivialised classical baseline.

Further modifications are required within PennyLane's framework to enable automatic backpropagation and to support differentiable workflows, essential for qutrit-based QML algorithms. For instance, since certain PennyLane functions do not allow the calculation of derivatives, differentiable tensors from the PyTorch library~\cite{pytorch} have to be used to preserve the workflow differentiability. While these limitations and the scaling down may potentially affect QAE performance, we have found their impact to be small compared with the reduction in circuit size, as shown in the Results section. The scaling to a bigger number of qutrits is left for future work.
\begin{figure*}[!t]
\includegraphics[width=17cm]{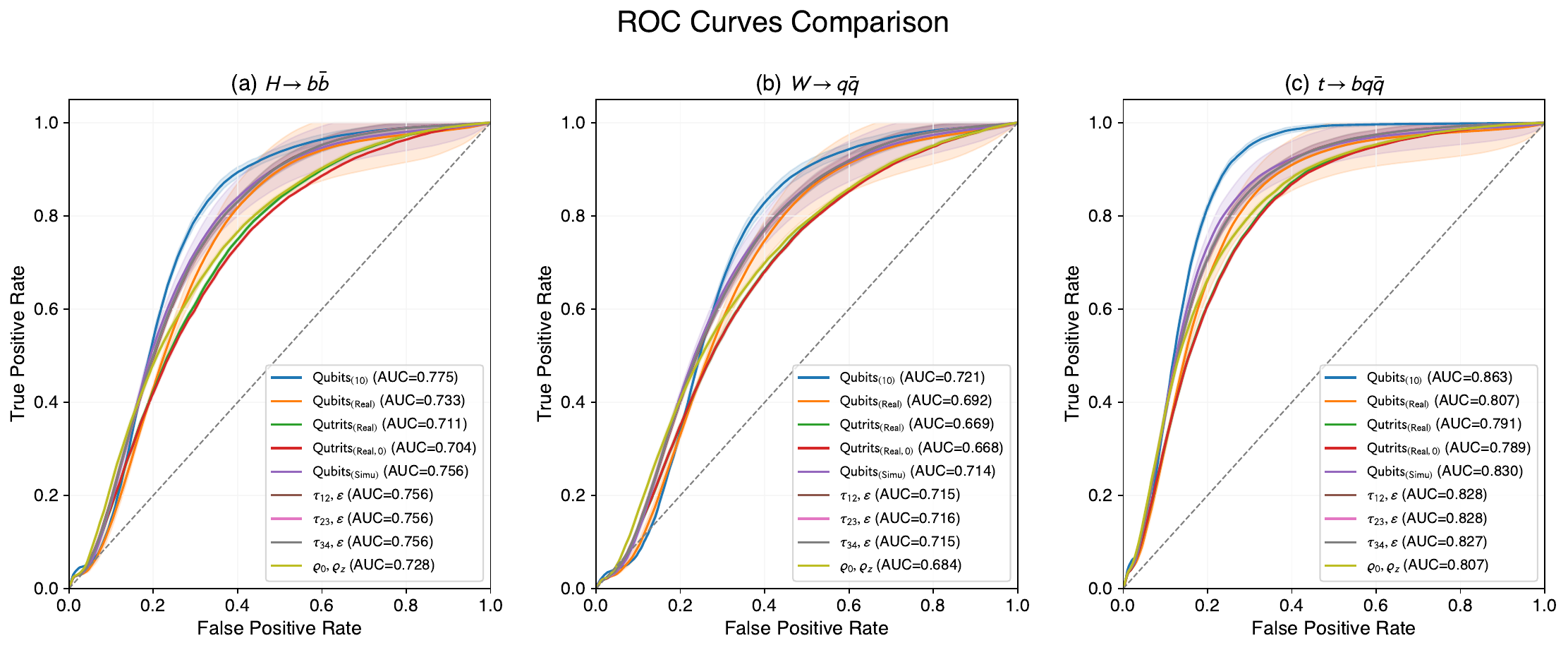}
\caption{\protect\footnotesize\label{fig:roc}\protect\justifying \textbf{ROC curves across the different models over 100 executions.} The solid lines represent the mean True Positive Rate (TPR) after interpolating multiple runs. The shaded bands indicate the $1\sigma$ standard deviation, accounting for the statistical stability and robustness of the models across different data partitions. For each signal (a-c), ROC curves are obtained by models with three trash qubits/qutrits and a latent space of one qubit/qutrit.}
\end{figure*}

\subsection*{\label{sec:Software}Simulation software}
In the model used as baseline, the circuit was simulated and optimised using the PennyLane devices \texttt{lightning.\allowbreak gpu} and \texttt{lightning.\allowbreak kokkos}, which are fast, high-performance simulators of state vectors and tensor networks written in C++. When it comes to replicating the structure presented in Ref.~\cite{1p1q}, \texttt{default.\allowbreak qubit}, a standard PennyLane quantum simulation device, is applied; it is written in Python and allows backpropagation-based derivative processes. 

Regarding the development of the QAE model based on qutrits, the PennyLane experimental device, \texttt{default.\allowbreak qutrit} for qutrit simulation, is used, which has native support for differentiable workflows through automatic backpropagation. PennyLane was chosen for the model development not only because it provides a stable simulation environment, but also because it enables an accurate comparison of results between the qubit-based model and the qutrit-based model.

\section*{\label{sec:Results}Results\protect}

\subsection*{\label{sec:level2}Reference model with qubits}
To verify the effectiveness of the 1P1Q method and the model's robustness, two approaches are adopted in Ref.~\cite{1p1q}. First, the encoding scheme is tested as part of a QAE model. Second, recorded and simulated experimental data from the CMS detector~\cite{aspenJets, jetClass} are used to demonstrate the practical application of the model to HEP events (see Ref.~\cite{1p1q} for more details), proving to be a highly effective and scalable quantum encoding strategy. To reproduce the results presented in Ref.~\cite{1p1q} and use them as a basis for benchmarking the performance of qubit- and qutrit-based models, we replicate the qubit-based QAE model.
\begin{table}[!h]
\centering
\caption{\label{tab:table1}\protect\justifying \textbf{AUC classical and qubit-based models scores.} Comparative AUC scores between the classical reference model (CAE Ref.), the reference model (QAE Ref.)\protect\cite{1p1q} and the replicated version (QAE Rep.) for a system with ten input qubits for the encoding and a latent space of two qubits. The reported (QAE Rep. 10) average AUC score and uncertainty ($\sigma$) correspond to 10 runs of the QAE Rep. model trained on simulated data.}
\vspace{5pt}
\begin{tabular*}{\columnwidth}{@{\extracolsep{\fill}}lccc}
\toprule
\textbf{Model} & $W \rightarrow q\bar{q}$ & $H \rightarrow b\bar{b}$ & $t \rightarrow bq\bar{q}$ \\
\midrule
CAE (Ref.) & $0.671$ & $0.739$ & $0.858$ \\
QAE (Ref.) & $0.715$ & $0.774$ & $0.872$ \\
QAE (Rep.) & $0.734$ & $0.776$ & $0.857$ \\
QAE (Rep. 10)   & $0.721 \pm 0.004$ & $0.775 \pm 0.004$ & $0.864 \pm 0.003$ \\

\bottomrule
\end{tabular*}
\label{tab:repl}
\end{table}

The results we obtain for the implementation using qubits, as well as the qubit-based results of the baseline model and the classical results presented in Ref.~\cite{1p1q}, are summarised in Table~\ref{tab:repl}, illustrating that the qubit-based QAE exhibits a behaviour in agreement with the reference model. Even without using a high-performance simulator (see details in the Simulation Software subsection of the Methods section), it shows a remarkable robustness in learning the fundamental substructure of the jet's characteristics. This makes this qubit-based QAE a suitable benchmark when developing a similar model based on qutrits and is therefore used as a reference for the remainder of this paper. Since Ref.~\cite{1p1q} presents only the results of a single run and does not provide any uncertainties, the average AUC scores over 10 runs, together with their uncertainties, are reported here in order to provide a reference for the full-size circuit. It should be noted that this increase in circuit size comes at a high computational cost, which is why we limit the sampling to 10 runs.
\begin{figure*}[!t]
\includegraphics[width=18cm]{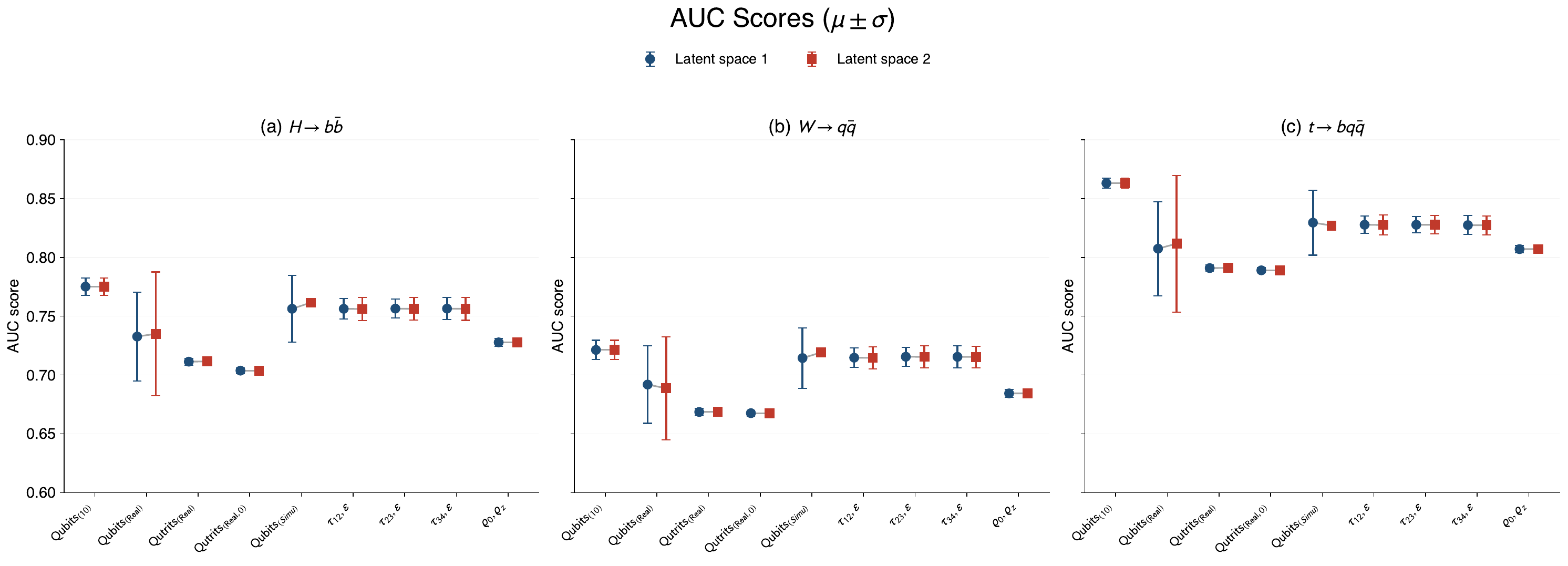}
\caption{\protect\footnotesize\label{fig:auc}\protect\justifying \textbf{Average AUC scores and uncertainty ($\sigma$) over 100 executions.} The scores obtained by the different QAE models across the various test signals are presented here. Notably, \textit{Qubits (10)} refers to the qubit-based full-size 10-particle replica, with AUC scores averaged over 10 executions. In addition, \textit{Qubits (10)} was tested only for latent space value $2$, while \textit{Qutrits (Real, 0)} was tested only for latent space value $1$, since in both cases the red and blue points correspond to the same repeated value.}
\end{figure*}
\subsection*{\label{sec:ExperimentalResults}Qutrit-based model results}
The qutrit-based model was tested under five different training parameter configurations, denoted A, B, C, D and E, corresponding to
\begin{itemize}
\item case A: $\phi_1 = \varrho_0$ and $\phi_2 = \varrho_z$
\item case B: $\phi_1 = \tau_{12}$ and $\phi_2 = \varepsilon$
\item case C: $\phi_1 = \tau_{23}$ and $\phi_2 = \varepsilon$
\item case D: $\phi_1 = \tau_{34}$ and $\phi_2 = \varepsilon$
\item case E: $\phi_1 = 0$ and $\phi_2 = 0$. 
\end{itemize}
All cases use the parameter configuration $\theta_1 = \varphi$ and $\theta_2 = \vartheta$. It is worth noting that in case E: $\Phi_1 = \Phi_2 = 0$, the input state carries the same information content as the standard qubit initialisation. However, this parameter choice does not reduce the physical system or the gate operations to the $d=2$ (qubit) case. The dynamics and quantum gates still act on the higher-dimensional Hilbert space governed by $SU(3)$ generators. Thus, this setting can be understood as embedding the qubit-like information into a specific subspace of the qutrit representation, rather than a dimensional reduction of the quantum model itself.

The evaluation of the qutrit-based model follows a structured three-step workflow. First, the model is trained using 10,000 background events, from either CMS proton-proton collision data or simulated CMS data. Second, 2,500 independent events from the same source as in the training are used as a validation set to prevent overfitting. Finally, the model's performance is evaluated by calculating anomaly scores for 10,000 unseen events of each signal type. These scores are compared against a test set of 10,000 previously unseen background events to perform a robust comparative analysis of the discrimination power.

The performance of the compression is evaluated through the quantum fidelity between the state of the trash qutrits, after the variational layers, and a reference state initialised to $|0\rangle$. 

\begin{figure*}[!t]
\includegraphics[width=17cm]{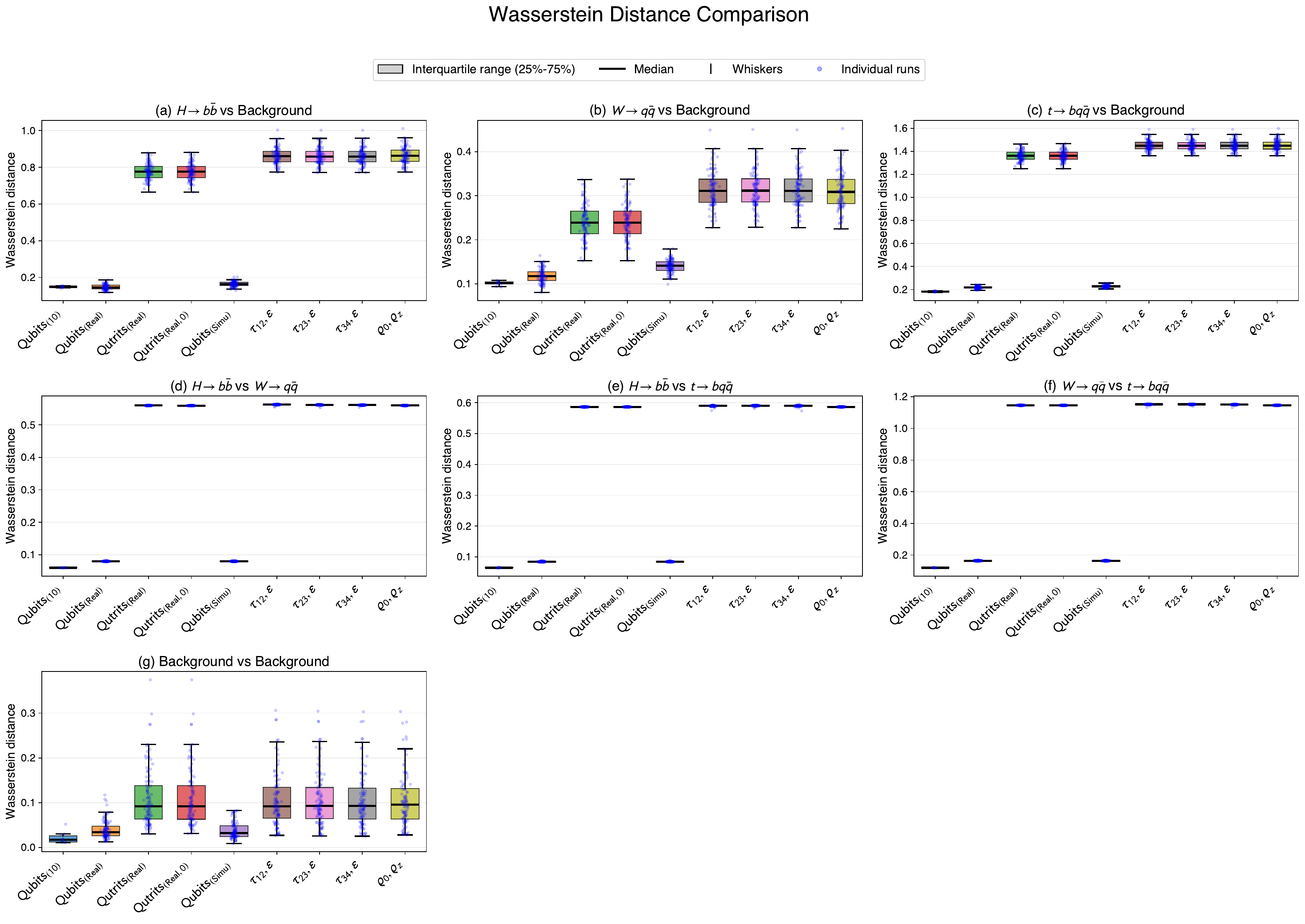}
\caption{\protect\footnotesize\label{fig:wassertein_boxplot}\protect\justifying \textbf{Distributions of Wasserstein distances ($W_1$) across the different models.}
Boxplots show the spread of $W_1$ distance values over 100 runs for comparisons including "signal vs. background" (a-c), "signal vs. signal" (d-f), and "background vs. background" (g). Median values and individual data points are highlighted, illustrating the statistical separation and variance for each condition.}
\end{figure*}

\begin{figure*}[!t]
\includegraphics[width=18cm]{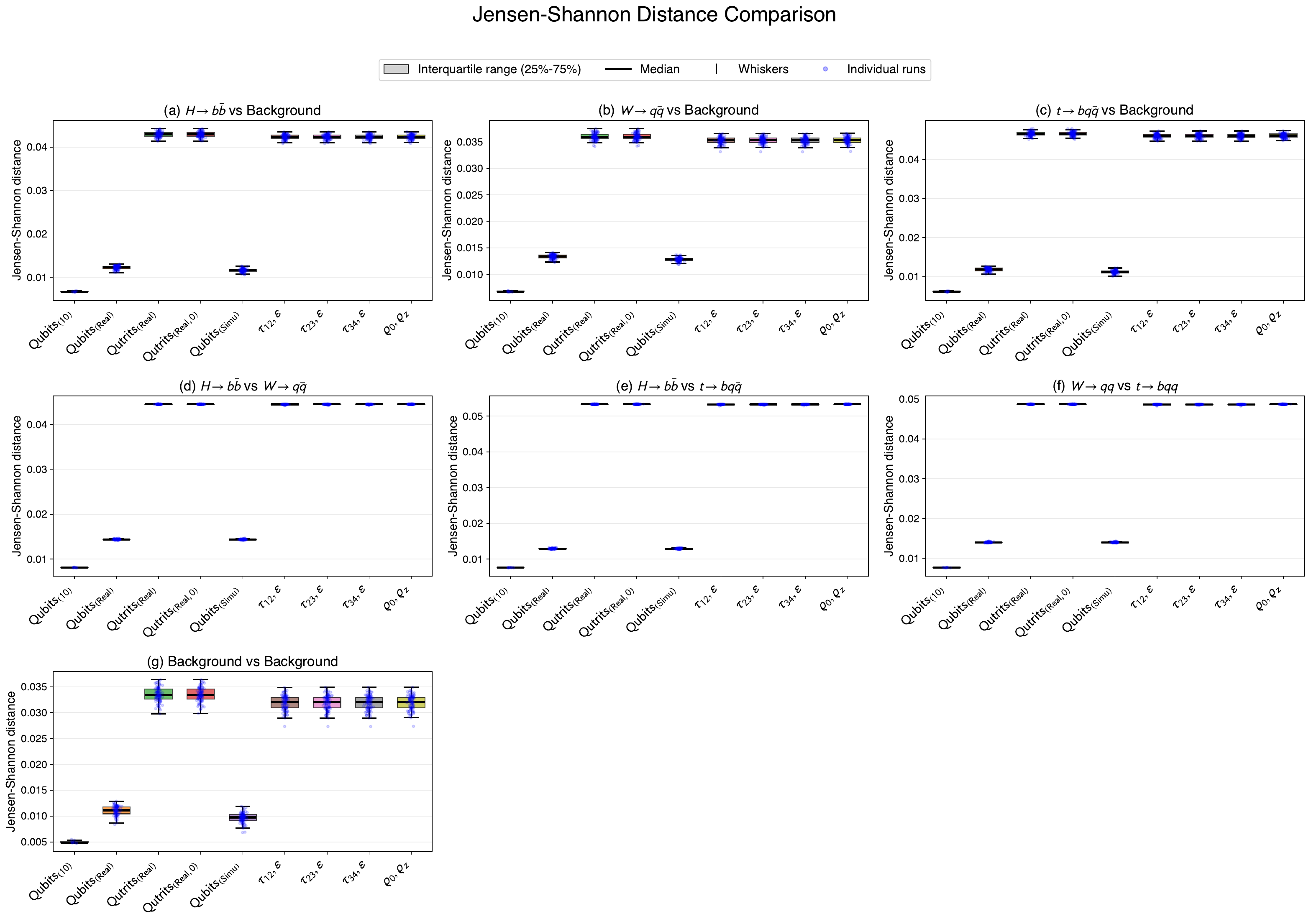}
\caption{\protect\footnotesize\label{fig:js_boxplot}\protect\justifying \textbf{Distributions of Jensen–Shannon distances (JS) across the different models.}
Boxplots show the spread of JS distance values over 100 runs for comparisons including "signal vs. background" (a-c), "signal vs. signal" (d-f), and "background vs. background" (g). Median values and individual data points are highlighted, illustrating the statistical separation and variance for each condition.}
\end{figure*}

\begin{figure*}[!t]
\includegraphics[width=18cm]{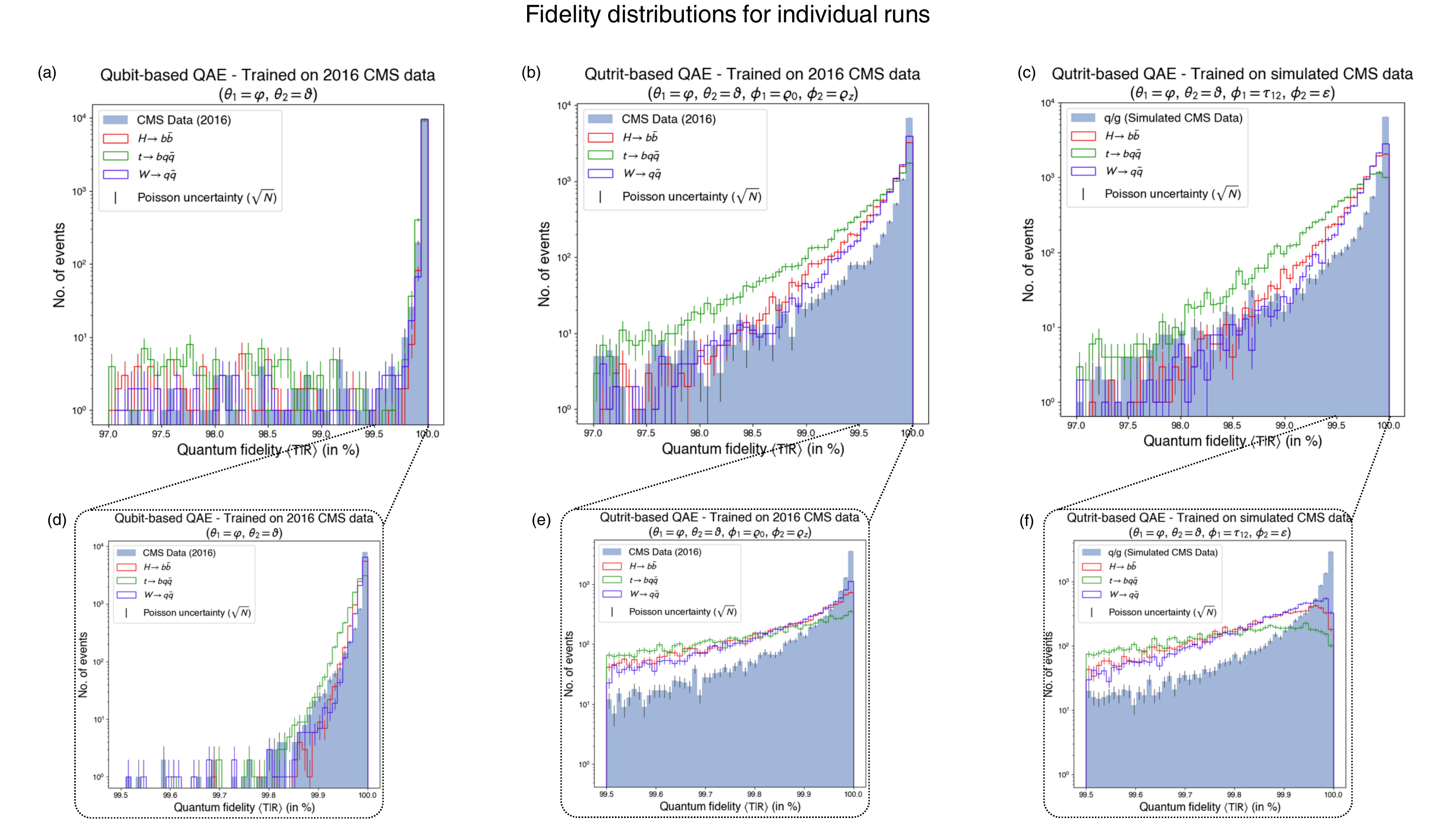}
\caption{\footnotesize\protect\justifying \textbf{Fidelity distributions for individual runs with statistical uncertainty.} For each model, using three trash qubits/qutrits and a latent space of one qubit/qutrit, the best run out of 100 was selected to plot the fidelity distribution with Poisson statistical uncertainty ($\sqrt{N}$), where $N$ denotes the number of events in each bin. Additionally, for each of the three cases (a)/(d), (b)/(e), and (c)/(f), the distributions are shown in the fidelity ranges of 97-100\% and 99.5-100\%.}
\label{fig:auc_plus}
\end{figure*}

The AUC scores, ROC Curves, Wasserstein distances and Jensen-Shannon distances are evaluated across 100 independent runs. The average values and their corresponding uncertainties are then calculated for two scenarios: the one with the model trained on 2016 CMS data, `Real', and the one with it trained on simulated data, `Simu'. The terms `Qutrits(Real, 0)' and `Qutrits(Real)' correspond to cases E and A, respectively, when the model is trained on 2016 CMS data. When neither \textit{Real} nor \textit{Simu} are specified, it always refers to models trained on simulated CMS data. These quantities are evaluated for all possible encoding cases, A-E, as shown in Fig.~\ref{fig:roc}, Fig.~\ref{fig:auc},  Fig.~\ref{fig:wassertein_boxplot}, and Fig.~\ref{fig:js_boxplot}. The best-performing run, over 100, is selected for each model to analyse its fidelity distribution, and the resulting quantum fidelity distributions are presented in Fig.~\ref{fig:auc_plus} for the fidelity range of $97\text{–}100\%$, and $99.5\text{–}100\%$. For both ranges, the associated Poisson statistical uncertainty ($\sqrt{N}$), where $N$ denotes the number of events in each bin, is also shown. 

Each run is trained and tested on a different data sample drawn from the original background and signal datasets. The True Positive Rate (TPR) and the False Positive Rate (FPR) are also calculated and represented as ROC curves, as shown in Fig.~\ref{fig:roc}. The solid lines represent the mean performance, while the shaded regions indicate the $1\sigma$ statistical uncertainty, derived from the standard deviation across all runs. These ROC curves confirm that all models maintain a high signal efficiency (TPR) even at low levels of background acceptance (FPR), which is crucial for anomaly detection in high-luminosity environments. It should be noted that, for this circuit size and a latent space of one qutrit/qubit, the qubit-based model appears to exhibit a higher level of uncertainty. This level decreases when the latent space is increased by one, but it remains consistently higher than that of the qutrit-based model. This is reasonable, since qutrits have access to a larger set of variables and greater descriptive power, leading to lower levels of uncertainty and higher stability.

In terms of fidelity distributions, and after imposing a threshold to ensure high-fidelity ranges, the qutrit-based model shows a clearer distinction among the relevant physical signals ($W \rightarrow q\bar{q}$, $H \rightarrow b\bar{b}$, and $t \rightarrow bq\bar{q}$), particularly in the fidelity range of $97\text{–}100\%$, thereby hinting at a greater expressive power of qutrit systems. In the more restrictive fidelity range of $99.5\text{–}100\%$, the distinction among the signals becomes visually less clear, although for fidelities close to $100\%$ the separation between fidelity distributions of signals and background remains quite evident, and the improved distinction among signals compared with the background is still indisputable. 

To be able to quantify the discrimination capability of the two models between signal fidelity distributions, and between background and signal fidelity distributions, two different metrics are used: the Jensen–Shannon distance ($\sqrt{JS}$) and the Wasserstein distance ($W_1$) over 100 executions. The results are summarised using boxplots in Fig.~\ref{fig:wassertein_boxplot} and Fig.~\ref{fig:js_boxplot}. Rather than reporting point estimates, the boxplots illustrate the median values along with the interquartile range (IQR) and statistical dispersion, capturing the variability in each model’s performance.
\begin{table*}[!t]
\centering
\caption{\label{tab:3}\footnotesize \textbf{Jet Energy Scale test modifying $p_T$ and }${P_{T,\text{jet}}}$\textbf{ by 3\% only in the training data.} Average Wasserstein distances and their uncertainty ($\sigma$), average Jensen–Shannon distances ($\sqrt{\textrm{JS}}$) and their uncertainty ($\sigma$), and average AUC scores and their uncertainty ($\sigma$) over 100 executions. (C) The qutrit model was trained using the parameters $\theta_1 = \varphi$, $\theta_2 = \vartheta$, $\phi_1 = \tau_{23}$, and $\phi_2 = E$.}
\vspace{5pt}
\begin{tabular*}{\textwidth}{@{\extracolsep{\fill}}lccc}
\toprule
\multicolumn{4}{c}{\textbf{Jensen--Shannon distances}} \\ 
\midrule
\textbf{F. Distributions} & $H - B$ & $t - B$ & $H - t$\\
\midrule
Qutrits S (C) ($\pm 3\% \,\,p_T$ | ${p_{T,\text{jet}}}$) & $0.04236 \pm 0.0005$ & $0.0460 \pm 0.0005$ & $0.05329 \pm 0.00004$ \\
Qutrits S (C) ($+ 3\% \,\,p_T$   | ${p_{T,\text{jet}}}$) & $0.0423  \pm 0.0005$ & $0.0460 \pm 0.0006$ & $0.05329 \pm 0.00006$ \\
Qutrits S (C) ($- 3\% \,\,p_T$   | ${p_{T,\text{jet}}}$) & $0.0423  \pm 0.0005$ & $0.0460 \pm 0.0005$ & $0.05329 \pm 0.00004$ \\
\midrule
\multicolumn{4}{c}{\textbf{Wasserstein distances}} \\ 
\midrule
\textbf{F. Distributions} & $H - B$ & $t - B$ & $H - t$\\
\midrule
Qutrits S (C) ($\pm 3\% \,\,p_T$ | ${p_{T,\text{jet}}}$) & $0.86 \pm 0.04$ & $1.45 \pm 0.04$ & $0.5904 \pm 0.0011$ \\
Qutrits S (C) ($+ 3\% \,\,p_T$   | ${p_{T,\text{jet}}}$) & $0.86 \pm 0.04$ & $1.45 \pm 0.04$ & $0.5903 \pm 0.0021$ \\
Qutrits S (C) ($- 3\% \,\,p_T$   | ${p_{T,\text{jet}}}$) & $0.86 \pm 0.04$ & $1.45 \pm 0.04$ & $0.5903 \pm 0.0013$ \\
\midrule
\multicolumn{4}{c}{\textbf{AUC Scores}} \\ 
\midrule
\textbf{Model} & $H \rightarrow b\bar{b}$ & $t \rightarrow bq\bar{q}$ & $-$ \\
\midrule
Qutrits S (C) ($\pm 3\% \,\,p_T$ | ${p_{T,\text{jet}}}$) & $0.757 \pm 0.007$ & $0.828 \pm 0.005$ & $-$ \\
Qutrits S (C) ($+ 3\% \,\,p_T$   | ${p_{T,\text{jet}}}$) & $0.756 \pm 0.009$ & $0.828  \pm 0.007$  & $-$ \\
Qutrits S (C) ($- 3\% \,\,p_T$   | ${p_{T,\text{jet}}}$) & $0.757 \pm 0.008$ & $0.828  \pm 0.007$ & $-$ \\
\bottomrule

\end{tabular*}
\end{table*}

For both metrics, qutrit-based models ultimately exhibit larger distances, as expected, not only between the signals themselves but also between signal and background, with the Wasserstein distance attaining the highest discrepancies. 
It is observed that the baseline background–background ($B$–$B$) Jensen-Shannon distance for the qutrit model is noticeably larger, indicating a greater intrinsic variance attributable to the geometry of the expanded $3^N$ Hilbert space. However, this limitation does not extend to the Wasserstein distance. The Jensen-Shannon distance provides a bounded ($[0,1]$), information-theoretic measure of overlap and shape deformation. The Wasserstein metric assesses the optimal transport cost along the horizontal fidelity axis, rather than focusing on vertical probability density overlaps, thereby rendering it inherently more robust to the structural baseline noise introduced by the expanded Hilbert space. Using both metrics ensures that the model's expressive power is evaluated both in terms of global shift in fidelity space ($W_1$) and local structural dissimilarity ($JS$). In the qutrit model, the signal–background Wasserstein distance remains substantially higher than its corresponding $B$–$B$ baseline. This distinction is crucial, as it demonstrates that the qutrit framework achieves genuine physical state separation, rather than merely amplifying statistical variance. 

Larger distances were also observed in models trained on simulated data, particularly for parameter combinations involving the $\tau$-ratios. This behaviour can be attributed to the idealised nature of the simulated dataset (JetClass), which accounts for detector effects in a simplified manner. Such conditions simplify the learning task and may favour simpler model architectures. 

In addition, models trained on simulation can exploit generator-specific correlations (e.g., from Pythia) that are not fully represented in real data. Furthermore, the enhancement in performance when using $N$-subjettiness ratios ($\tau_{12}$, $\tau_{23}$, $\tau_{34}$) compared to the impact parameters ($d_0$, $d_z$) stems from the physical information they encode: jet substructure versus displaced vertices, where $\tau$-ratios provide a more powerful discriminating signal. Impact parameters measure the displacement of a particle’s trajectory from the primary interaction point and are highly effective for identifying b-jets. They essentially act as binary indicators and provide limited information about the jet’s energy distribution. On the other hand, the $\tau$-ratios describe the topology of the radiation. For instance, $\tau_{12}$ tells the model if a jet looks more like a single blob of energy (background QCD) or two distinct subjets, e.g., a boosted W or H boson decaying into two quarks. Concerning the values obtained for both metrics, and taking as a reference the distance between background and background fidelity distributions ($B-B$), which is expected to be comparatively low, it appears that the Wasserstein distance constitutes a more appropriate metric for assessing the model’s capacity to discriminate between distinct signals and for quantifying the similarity between fidelity distributions, owing to its ability to capture the underlying structure and geometry of the data distributions. For case E, in which the additional angular parameters $\phi_1$ and $\phi_2$ are set to zero, it appears that the distance metrics are not affected. This points to the increased expressive power, understood in terms of discrimination between signals and background and the distance between fidelity distributions, being a natural feature of qutrits. Moreover, when further examining case E, a clear decrease in performance in terms of AUC scores is observed compared to scenarios (A-D), which is expected since we are introducing less information about the constituents.

In order to study how the size of the latent space affects performance, all scenarios except case E were re-evaluated by increasing the latent space from one to two qubits/qutrits. Since the differences in terms of distance metrics were almost imperceptible for qutrits and even smaller for qubits, only the results in terms of AUC scores are included and summarised in Fig.~\ref{fig:auc}. As can be seen, increasing the latent space does not appear to produce a remarkable improvement. For qubits, in some cases, a larger latent space seems to slightly improve the AUC score and reduce uncertainty, thereby stabilising the model. However, for qutrits, no improvement is observed either in AUC score performance or in the level of uncertainty. This suggests that, for this specific circuit size, qutrit-based models are less sensitive to increases in the latent space.
Moreover, when the model is trained on simulated data for both qutrits and qubits, and when the qutrit-based model includes information about the $N$-subjettiness parameters, it achieves higher AUC scores than when it is trained on CMS 2016 data or when impact parameters are used, as expected. 

Although it was necessary to reduce the circuit size due to PennyLane’s limitations in simulating qutrits and the associated memory consumption, when considering the results for the full 19-qubit circuits, the performance degradation observed in the scaled-down version does not appear to be particularly high. This is especially noteworthy given that the circuit size was reduced from 19 qubits to 8 qubits/qutrits, and that the encoding was reduced from 10 particles per iteration and 10 constituents per jet to only 4 particles per iteration and 4 constituents per jet. The most significant improvement was observed for the distance metrics. 

In order to evaluate the impact of restricting the input representation to four particles, the transverse momentum ($p_T$) distribution of the leading constituents across the simulated CMS dataset has been analysed. The four leading constituents have systematically captured the vast majority of the jet’s energy flow, averaging approximately $80\text{–}100$ GeV compared to the $\sim 50\text{–}60$ GeV average across the full constituent spectrum, as well as the characteristic prong-like topology of the signals that is essential for effective discrimination. Although the omission of additional soft constituents has removed some information on wide-angle radiation, the high-fidelity reconstruction of these four leading particles appears sufficient for the QAE to distinguish the background norm from anomalous signal structures.

To assess the model’s sensitivity to the Jet Energy Scale (JES), three tests were performed by modifying $p_T$ and ${p_{T,\text{jet}}}$ in the training data by $3\%$. We took this number as a conservative estimate of the overall systematic uncertainty due to JES using a CMS calibration~\cite{CMS:2016lmd}. In the first case, $p_T$ and ${p_{T,\text{jet}}}$ were always increased by $3\%$; in the second case, they were always decreased by $3\%$; and in the third case, the $3\%$ increase or decrease was applied randomly. For all cases, the Wasserstein and Jensen–Shannon distances, as well as the AUC scores, were calculated, as shown in Table~\ref{tab:3}. Taking these and the other results into consideration, it appears that slight fluctuations in $p_T$ do not significantly affect the performance of the qutrit-based model.

In light of our results, the $t \to b q \bar{q}$ decay is identified by the QAE as the most anomalous signal by all models. This is consistent with the expected jet topology: at LHC energies, electroweak-scale resonances such as the top quark, $W/Z$ bosons, and the Higgs boson are frequently produced in a boosted regime, i.e. their energy significantly exceeds their mass, and their decay products are collimated into a single large-radius jet. While QCD jets typically exhibit a single-prong core with diffuse soft radiation, boosted  $W/Z$ and Higgs bosons yield a two-prong structure, and top quarks produce a more complex three-prong configuration~\cite{JETS,BOOSTED}. The richer substructure and higher particle multiplicity in top-initiated jets create larger deviations from the QCD-like background, explaining why the QAE detects $t \to b q \bar{q}$ events as more anomalous, followed by $H \to b \bar{b}$ and then $W$-boson decays. Accordingly, the inclusion of the ratios $\tau_{12}$, $\tau_{23}$ and $\tau_{34}$ together with the jet energy enhances the discrimination power for $t \to b q \bar{q}$ events, leading models B, C and D to achieve higher AUC values and distance metrics than model A, which uses the longitudinal and transverse impact parameters as features.

\section*{\label{sec:Discussions}Discussion\protect}

Despite the limitations of qutrit-based simulators, a level of performance almost equivalent to the qubit-based model is achieved in terms of AUC scores, especially when training is performed on simulated data. Thus, a comparable ability to distinguish anomalous events from QCD jets without prior anomaly labelling is achieved. Also, it appears that a variation in the latent space affects only qubit-based models, slightly improving AUC scores and reducing uncertainty levels. For qutrit-based models, no effects are observed in either the AUC scores or the uncertainty levels, which are generally lower than those of qubit-based models, suggesting that their higher descriptive power provides greater stability in the results. In terms of distances between fidelity distributions, as well as the capacity to discriminate between different physical signals, all qutrit-based models achieved higher values than qubit-based models, indicating a greater capacity to discern between the three types of signals and therefore a greater ability to detect substructural differences between them.

Furthermore, Majorana encoding makes it possible to encode more data in the same circuit size compared to the qubit-based model, enriching the model's expressive power and exploiting its greater compactness to reduce the circuit's depth. Even when the additional parameters provided by the Majorana encoding are set to zero, no significant decrease is observed in terms of AUC scores, nor is any detectable decrease found in the distance-based metrics. Since the two additional parameters in the Majorana encoding appear as phases, it seems reasonable that, even though they incorporate more information about the constituents or jets, they do not provide sufficient improvement to outperform qubit-based models for circuits of this size. 

This makes qutrit-based models an interesting candidate for further development, especially as advances in qutrit hardware continue, which makes the implementation of more complex tests possible,  rendering them a possible alternative for detecting subtle differences or anomalous data at the LHC, which could play a fundamental role in the discovery of physics beyond the Standard Model. Additionally, the Majorana encoding opens the door for the implementation of other LHC challenges in qutrit-based models. 

\section*{Conclusions\protect}
A qutrit-based model for anomaly detection in CMS experiment data has been developed, and Majorana encoding for qutrit representation on unit spheres has been proven to be an effective way to represent the information on a unitary sphere for qutrit systems. Generalised gates have been added according to the new ternary paradigm, such as rotation gates and SWAP gates, which have been shown to give robust results. Despite the limitations encountered in PennyLane's ability to simulate qutrit circuits, in terms of memory consumption, a performance comparable to the qubit-based model has been achieved, and our model, in terms of distance metrics, has stood out for its greater ability to discern between signals.

Regarding future research, further tests should be carried out using data from other LHC experiments to evaluate how the new encoding and model behave with different datasets and Beyond the Standard Model scenarios. In addition, different encodings and libraries should be studied and tested to determine the most suitable configurations for qutrit-based analyses of LHC data.

Concerning the hardware implementation of qudit systems, recent studies, such as the one presented in Ref.~\cite{trappedIon}, where a quantum algorithm was successfully implemented with a trapped ion qudit, obtaining competitive results, make the hardware implementation of qudit-based models seem like a certainty in the intermediate future. Additionally, the break-even point for Quantum Error Correction of qudit quantum memories, at which the lifetime of a qudit exceeds the lifetime of the constituents of the system, has been beaten in Ref.~\cite{evenBreak}, enhancing the usability conception of qudits in the long run and making studies like this one only the beginning of further developments.

The use of qutrit mixed states, using Majorana's generalised representation for mixed states~\cite{mixed_states}, as well as higher-level systems such as ququarts~\cite{ququart}, which can potentially provide greater expressive power and thus improve the performance of our model for anomaly detection, is left for future work. It should be noted that these performance improvements come at an increased computational cost and are not a straightforward implementation within the available simulators.

\section*{Data availability\protect}

All the raw datasets used in this work for model training and evaluation are available at \href{https://www.fdr.uni-hamburg.de/record/16505}{{\color{blue}https://www.fdr.uni-hamburg.de/record/16505}} and \href{https://zenodo.org/records/6619768}{{\color{blue}https://zenodo.org/records/6619768}}. The raw-result datasets, together with the processed results in tabular format, are available at \href{https://doi.org/10.6084/m9.figshare.33297300}{{\color{blue}https://doi.org/10.6084/m9.figshare.33297300}} and \href{https://doi.org/10.6084/m9.figshare.33296889}{{\color{blue}https://doi.org/10.6084/m9.figshare.33296889}}. All data produced or used in this work are open source.

\section*{Code availability\protect}

All the code used in this work is open source and available at \href{https://github.com/MirandaCarou/Qutrits-for-physics-at-LHC}{{\color{blue} https://github.com/MirandaCarou/Qutrits-for-physics-at-LHC}}.  To facilitate accessibility and reuse, we provide both Jupyter notebooks, intended for didactic and illustrative purposes, and the complete implementation in the form of standard Python scripts. These scripts correspond to the versions used to execute all experiments and can be employed to reproduce the results reported in the manuscript.

\section*{Author contributions}
M.C.L. developed the software, performed the simulations, and identified the specific algorithmic framework used in the final implementation. V.C. contributed to the unitary representation and the interpretation of results. M.L.M. proposed the use case and provided technical supervision. M.L.M. and V.C. co-supervised the study. All authors contributed to regular discussions, interpretation of results and the final manuscript.
\section{Competing interests}
Miranda Carou Laiño, Veronika Chobanova and Miriam Lucio Martínez declare no competing interests.

\begin{acknowledgments}
M. Carou and V. Chobanova acknowledge support from the InTalent program (Inditex–UDC). V. Chobanova is a Ramón y Cajal researcher, RYC2021-031287-I, funded by NextGenerationEU, the Spanish State Research Agency, and the Spanish Ministry of Science, Innovation and Universities projects PID2022-139514NB-C32 and PCI2023-145957-2. V. Chobanova also acknowledges funding from Xunta de Galicia (ED431F 2023/40). M. Lucio is a Ramón y Cajal researcher, RYC2023-043882-I. M. Carou acknowledge support from the Ministry of Economic Affairs and Digital Transformation of the Spanish Government through the QUANTUM ENIA project call - QUANTUM SPAIN project, and from the European Union through the Recovery, Transformation and Resilience Plan - NextGenerationEU within the framework of the Digital Spain 2026 Agenda. We also acknowledge A. Bal, M. Klute, B. Maier, M. Oughton, E. Pezone, and M. Spannowski for helpful discussions; A. Chavert Sancho for helpful discussions about the metrics; and B. Pérez Dominguez for technical support with high-performance computing environments, as well as the Galician Supercomputing Centre (CESGA) for providing access to computing resources on the high-performance computer FinisTerrae III.
\end{acknowledgments}

\appendix
\section{\label{app:features} Feature encoding analysis}
To determine the optimal input representation without introducing selection bias or data leakage, a systematic feature selection study was conducted on a dedicated set from the entire JetClass dataset, partitioned into $10{,}000$ training, $2{,}500$ validation, and $10{,}000$ unused testing events for background, alongside dedicated signal samples. Crucially, this set of data was used exclusively for exploring input representations and was kept completely disjoint from the data employed in the final evaluations. 

To avoid prohibitive computational overhead while performing an unbiased search, feature selection was restricted to unique four-variable combinations rather than ordered permutations. Evaluating all possible ordering permutations would lead to a total of $3{,}024$ possible combinations, which is computationally intractable for quantum-circuit simulations. Instead, all $C(9,4) = 126$ unique combinations of four input variables were systematically evaluated on this dedicated validation set. Guided by the results of this 126-combination study, as well as physical motivation (since spatial coordinates are standard starting points in jet-tagging baselines), $\eta$ and $\phi$ were held fixed in the first two quantum registers for the final model architecture. The remaining two registers were used to encode complementary kinematic and substructure features, while physically redundant pairings (such as energy $E$ with invariant mass $m$) were systematically excluded. The resulting optimal combination, hereafter used as the baseline encoding for qutrits, is the one in which $\theta_1 = \vartheta$ and $\theta_2 = \varphi$ remain fixed (as in Ref.~\cite{1p1q}), while $\phi_1 \in \{\tau_{12}, \tau_{23}, \tau_{34}\}$ with $\phi_2 = \varepsilon$, and $\phi_1 = \varrho_0$ with $\phi_2 = \varrho_z$ yielding the highest performance. 

In the final results presented in the main manuscript, the decision to fix the angles $\theta_1 = \vartheta$ and $\theta_2 = \varphi$ was also made to align directly with the spatial and kinematic coordinate mapping, since the transverse momentum $p_T$, pseudorapidity $\eta$, and azimuthal angle $\phi$ fully determine the particle's kinematics. Mapping these variables to the primary rotation angles ensures that the complete physical momentum and directional information of each particle is compactly preserved. Thus, the impact of the additional qutrit degrees of freedom (the phases $\phi_1, \phi_2$) can be systematically isolated and evaluated without distorting the underlying detector-level information. Furthermore, comparing alternative parameter assignments confirmed that preserving this full physical mapping in $\theta_1$ and $\theta_2$ yields the optimal classification performance, thus validating this configuration as the richest in information. The correlation matrix, which illustrates the interdependence between global jet features and individual particle-level constituents in the JetClass dataset, is shown in Supplementary Fig.~\ref{fig:corr}.

\begin{figure}[h]
\includegraphics[width=8cm]{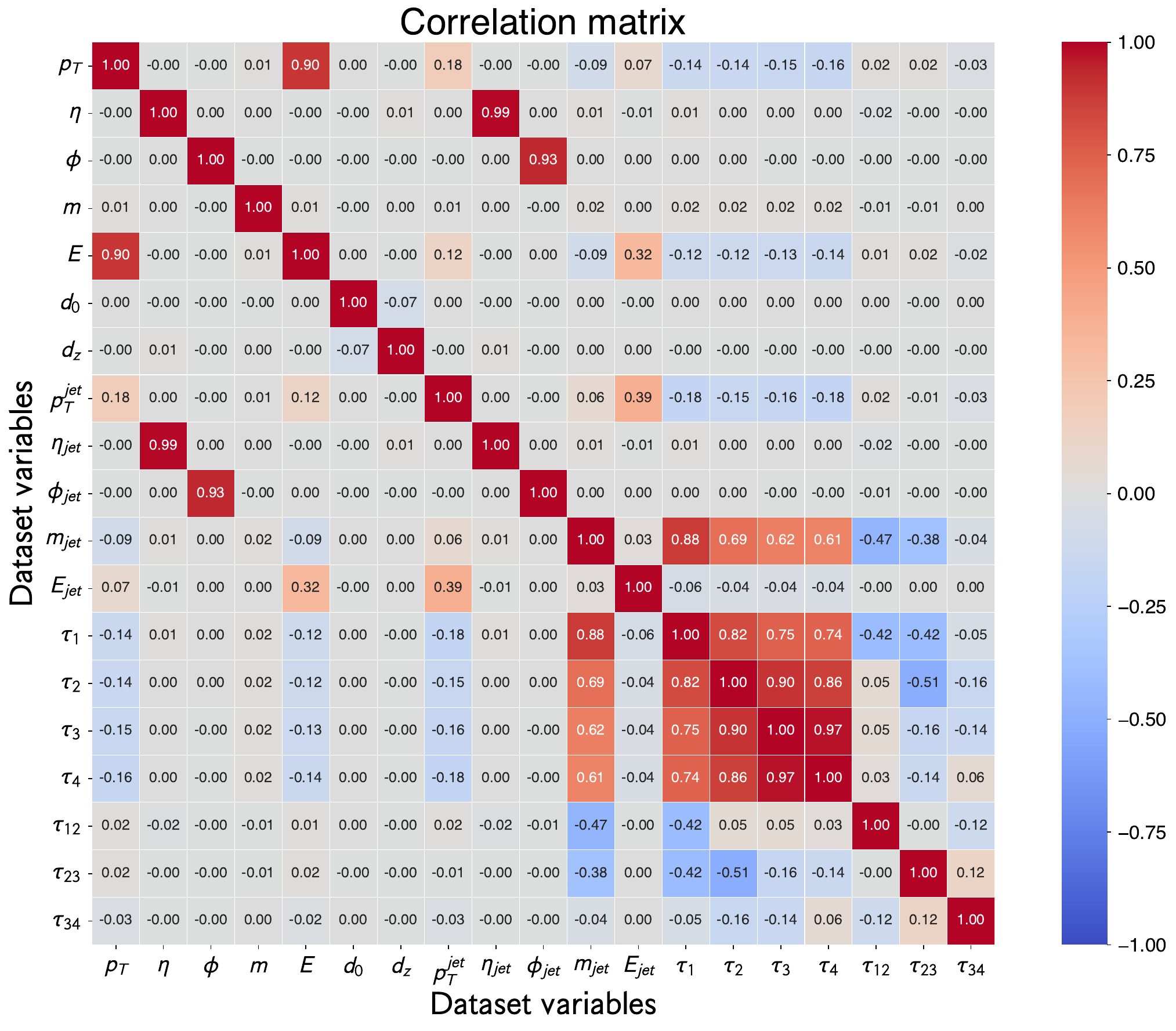}
\caption{\protect\footnotesize \protect\justifying \textbf{Correlation matrix.} The interdependence between global jet features and individual particle-level constituents in the CMS simulated samples is represented such that reddish hues indicate stronger intercorrelation, whereas bluish hues indicate weaker intercorrelation.}
\label{fig:corr}
\end{figure}

\section{\label{app:proof} Majorana general pure state derivation }
The trajectory of a point on the Bloch sphere can describe the evolution of a two-level quantum system. Similarly, in 1932, Italian theoretical physicist Ettore Majorana proposed a method to describe high-dimensional spaces with multiple points called \textit{stars} on the Bloch sphere, often referred to as the Majorana stellar representation (MSR) \cite{Majorana1932}. 

The Majorana stellar representation provides a geometric framework for visualising pure states of a $d$-level system (qudits) as a constellation of $d-1$ points, or \textit{stars}, on the surface of the Bloch sphere. For the specific case of a qutrit ($d=3$), a pure state is represented by two points on the unit sphere. Physically, this corresponds to expressing the spin-$1$ state as the symmetric tensor product of two constituent spin-$1/2$ states (spinors).

Let the two constituent spin-$1/2$ states be defined in the computational basis as:
\begin{flalign}
    &\ket{\psi_1} = \alpha\ket{0} + \beta\ket{1},\\
    &\ket{\psi_2} = \gamma\ket{0} + \delta\ket{1}
\end{flalign}
where the complex coefficients satisfy the normalization conditions $|\alpha|^2 + |\beta|^2 = 1$ and $|\gamma|^2 + |\delta|^2 = 1$. By mapping these states onto the Bloch sphere, they can be parameterised using the polar angles $\theta_i \in [0, \pi]$ and azimuthal angles $\phi_i \in [0, 2\pi]$ as follows:
\begin{eqnarray}
    &\ket{\psi_1} = \cos\left(\frac{\theta_1}{2}\right)\ket{0} + e^{i\phi_1} \sin\left(\frac{\theta_1}{2}\right)\ket{1}\\\nonumber
    &\ket{\psi_2} = \cos\left(\frac{\theta_2}{2}\right)\ket{0} + e^{i\phi_2} \sin\left(\frac{\theta_2}{2}\right)\ket{1}
\end{eqnarray}
The resulting qutrit state is then constructed via the projection onto the symmetric subspace of the tensor product of these two spinors:
\begin{flalign}
\ket{\psi_1} \otimes \ket{\psi_2} &= [\cos\left(\tfrac{\theta_1}{2}\right)\ket{0} + e^{i\phi_1} \sin\left(\tfrac{\theta_1}{2}\right)\ket{1}] \otimes \\[0.5em]  
&\otimes [\cos\left(\tfrac{\theta_2}{2}\right)\ket{0} + e^{i\phi_2} \sin\left(\tfrac{\theta_2}{2}\right)]\ket{1} = \\[0.5em]
&= \cos\left(\tfrac{\theta_1}{2}\right) \cos\left(\tfrac{\theta_2}{2}\right) \ket{00} + \\[0.5em]
&+e^{i\phi_2} \cos\left(\tfrac{\theta_1}{2}\right) \sin\left(\tfrac{\theta_2}{2}\right)\ket{01} +\\[0.5em]
&+ e^{i\phi_1} \sin\left(\tfrac{\theta_1}{2}\right) \cos\left(\tfrac{\theta_2}{2}\right) \ket{10}+ \\[0.5em]
&+ e^{i(\phi_1+\phi_2)} \sin\left(\tfrac{\theta_1}{2}\right) \sin\left(\tfrac{\theta_2}{2}\right) \ket{11} =\\[0.5em]
&=\ket{\Psi}
\end{flalign}
To obtain the qutrit state, we project this result onto the triplet (symmetric) subspace. In the total spin basis $\{ \ket{S,m} \}$ with $S=1$, the mapping is defined by the Clebsch-Gordan coefficients: $\ket{1,1}=\ket{00}$, $\ket{1,-1}=\ket{11}$, and $\ket{1,0}=\frac{1}{\sqrt{2}}(\ket{01}+\ket{10})$. Thus, the unnormalized qutrit state $\ket{\bar{\Psi}}$ is:
\begin{flalign}
\ket{\bar{\Psi}} &= \cos\left(\tfrac{\theta_1}{2}\right) \cos\left(\tfrac{\theta_2}{2}\right) \ket{1,1} + \\[0.5em]
&+\tfrac{1}{\sqrt{2}} [ e^{i\phi_2} \cos\left(\tfrac{\theta_1}{2}\right) \sin\left(\tfrac{\theta_2}{2}\right) +\\[0.5em]
&+ e^{i\phi_1} \sin\left(\tfrac{\theta_1}{2}\right) \cos\left(\tfrac{\theta_2}{2}\right) ] \,\,\ket{1,0} + \\[0.5em]
&+ e^{i(\phi_1+\phi_2)} \sin\left(\tfrac{\theta_1}{2}\right) \sin\left(\tfrac{\theta_2}{2}\right) \ket{1,-1}
\end{flalign}
By applying a global scaling factor of $\sqrt{2}$ and expressing the state in vector form, we obtain:
\begin{flalign}
& \ket{\bar{\Psi}} =  \nonumber\\
& = \begin{bmatrix}
\sqrt{2}\cos\left(\frac{\theta_1}{2}\right) \cos\left(\frac{\theta_2}{2}\right) \\[0.5em]
 e^{i\phi_2} \cos\left(\frac{\theta_1}{2}\right) \sin\left(\frac{\theta_2}{2}\right) + e^{i\phi_1} \sin\left(\frac{\theta_1}{2}\right) \cos\left(\frac{\theta_2}{2}\right)  \\[0.5em]
\sqrt{2}e^{i(\phi_1+\phi_2)} \sin\left(\frac{\theta_1}{2}\right) \sin\left(\frac{\theta_2}{2}\right)
\end{bmatrix} &
\end{flalign}
The norm of this state, $\mathcal{N}^2 = \left| \langle\Psi|\Psi\rangle \right|$, is evaluated using standard trigonometric identities, assuming $\theta_i, \phi_i \in \mathbb{R}$:
\begin{equation}
\frac{1}{2}[3 + \cos(\theta_1)\cos(\theta_2) + \sin(\theta_1)\sin(\theta_2)\cos(\phi_1 - \phi_2) ].
\end{equation}
To ensure the state is properly normalised, we define the normalisation factor $\Gamma = \mathcal{N}^{-1}$. Consequently, the physical qutrit state is given by $\ket{\Psi} = \Gamma \ket{\bar{\Psi}}$, where:
\begin{flalign}
&\Gamma = \nonumber\\
&\sqrt{2}[3 + \cos(\theta_1)\cos(\theta_2) + \sin(\theta_1)\sin(\theta_2)\cos(\phi_1 - \phi_2)]^{-\frac{1}{2}}.
\end{flalign}
\section{\label{app:Robust}Consistency check of the Majorana encoding implementation}
To validate the Majorana encoding and the generalised rotation gates, we compare the analytical framework with numerical simulations in PennyLane. This consistency check ensures the correct implementation of the unitary operators within the computational model. The aim of this appendix is not only to show that the operations in PennyLane work as expected from a mathematical perspective, but also to explain how the qutrit state can be obtained from four angles, how those angles can be recovered from the state, and how to operate with rotation gates.

Throughout this check, we transition between two equivalent representations of the qutrit Hilbert space depending on the context. The physical Majorana encoding is naturally described in the spin-1 basis, denoted by the magnetic quantum numbers $m \in \{-1, 0, 1\}$. However, for the numerical implementation and circuit operations within PennyLane, we adopt the standard computational basis $|n\rangle \in \{|0\rangle, |1\rangle, |2\rangle\}$. The mapping between these representations is defined as $|0\rangle \equiv |-1\rangle$, $|1\rangle \equiv |0\rangle$, and $|2\rangle \equiv |1\rangle$.
\subsubsection{\label{MajorabaRobust}Majorana encoding check}
Starting from a set of four arbitrarily chosen angles,
$\theta_1 = \frac{\pi}{2}$, $\theta_2 = \frac{\pi}{4}$, $\phi_1 = \frac{3\pi}{2}$, and $\phi_2 = \frac{4\pi}{3}$, we construct the qutrit state $\ket{\psi}$ by applying the Majorana encoding, according to Eq.~(6) of the Main manuscript, as a unitary operator.
\begin{equation*}
    \ket{\psi} = \begin{bmatrix}
\phantom{+}0.69 + 0.00\,i \\
-0.10 - 0.66\,i \\
-0.25 + 0.14\,i
\end{bmatrix}.
\end{equation*}
In the context of qutrit-based PennyLane circuits, the initial state is equivalent to the computational basis state $\ket{0}$, which is the basis state $\ket{-1}$ in the Majorana representation. To convert this state into a unitary operator, where the Majorana state coincides with the value of the first column and therefore allows this operator to be used to apply the initial encoding, the $QR$ decomposition method~\cite{QR} is applied. The $QR$ decomposition for a square matrix $A$ is given by
\begin{equation}
    A = QR
\end{equation}
where $Q$ is an orthogonal and unitary matrix ($Q^TQ = I$) and $R$ is an upper triangular matrix. 
This process gives rise to a unitary operator, with the initial state encoded in the first column,
\begin{equation}
U = Q = [ e_1, e_2, e_3].
    \label{eq:vector}
\end{equation}
Consequently, the vectors $e_{1,2,3}$ form an orthonormal set. They are defined in terms of the vectors $u_{1,2,3}$, which are the required system of orthogonal vectors, and the calculation of the sequence is known as Gram–Schmidt orthogonalization~\cite{smith_process}
\begin{equation}
    {e}_1 = \frac{{u}_1}{\|{u}_1\|}, \quad
{e}_2 = \frac{{u}_2}{\|{u}_2\|}, \quad
{e}_3 = \frac{{u}_3}{\|{u}_3\|}.
\label{eq:orthogonal}
\end{equation}
For our specific choice of angles, the unitary operator $U$ corresponds to
\begin{equation*}
   U =  \begin{bmatrix}
 -0.69 + 0.00i & -0.52 - 0.07i & \phantom{+}0.29 - 0.41i \\
  \phantom{+}0.10 + 0.66i & -0.10 - 0.21i & \phantom{+}0.58 + 0.41i \\
  \phantom{+}0.25 - 0.14i & -0.81 + 0.07i & -0.29 + 0.41i
\end{bmatrix}
\end{equation*}
which have to be applied to the initial states to perform the Majorana encoding.

The angles can be obtained by performing the reverse process. The Majorana qutrit state from Eq.~(5) of the Main manuscript is taken as a starting point, knowing that an arbitrary state of a qutrit in the computational basis $\{\ket{0}, \ket{1}, \ket{2}\}$ is given by:
\begin{eqnarray}
    \ket{\psi} = c_0\ket{0} + c_1\ket{1} + c_2\ket{2}
\end{eqnarray}
where $c_0$, $c_1$ and $c_2$ are complex probability amplitudes. The second-degree Majorana polynomial associated with the qutrit state satisfies
\begin{equation}
a_{2}\,\zeta^{2} + a_{1}\,\zeta^{1} + a_{0} = 0
\label{eq:poly}
\end{equation}
and is obtained by identifying the $s = 1$ basis, denoted by the state vectors $|sm\rangle_{\text{s}}  \ (m = -1, 0, 1)$, with the computational basis $|s+m\rangle_{\text{c}}$ as : $|sm\rangle_{\text{s}} \equiv|s+m\rangle_{\text{c}}$  (i.e., $c_0 = C_{-1}$, $c_1 = C_0$ and $c_2 = C_1$)~\cite{MISALVACION}. Therefore, the coefficients of the polynomial are obtained from
\begin{subequations}
\begin{eqnarray}
&a_{r} = (-1)^{r} \frac{C_{1-r}}{\sqrt{(2-r)!\,r!}} 
\label{eq:poly2}
\\\nonumber\\
&a_0 = \frac{c_2}{\sqrt{2}},\quad a_1 = -c_1,\quad a_2 = \frac{c_0}{\sqrt{2}}.
\label{eq:aleluya}
\end{eqnarray}
\label{eq:roots}
\end{subequations}
The two roots of the second-degree Majorana polynomial are:
\begin{subequations}
   \begin{eqnarray}
    &
    \zeta_i = e^{i\phi}\tan(\frac{\theta_{i}}{2})\quad i = 1,2\\\nonumber\\
    &\theta_i = 2\arctan|\zeta| \quad \phi_i= \arg(\zeta).
    \label{eq:angles}
\end{eqnarray} 
\end{subequations}
Taking $\theta_i$ and $\phi_i$ as shown in Eq.~(\ref{eq:angles}), we find that the angles extracted match the initial input parameters. This process is repeated using a simplified version of the model's circuit to verify that the results are consistent, thus confirming the correct implementation of the Majorana encoding within the model.
\subsubsection{Rotation gates}
To apply the rotation matrices $\Sigma_i$ to the circuit as rotations of the two points of the Majorana representation, the exponentiation according to Eq.~(4) of the Main manuscript must be performed. Taking as numerical values the angles $\varepsilon = \pi/2$, $\varphi = \pi/4$ and $\omega = 4\pi/3$, the resulting matrices with a precision of one significant digit are
\begin{eqnarray}
R_x(\varepsilon) &=&
\begin{bmatrix}
\phantom{+}0.5\phantom{i} & \phantom{+}0.7 i & -0.5 \\
\phantom{+}0.7 i & \phantom{+}0 & \phantom{+}0.7 i \\
-0.5\phantom{i} & \phantom{+}0.7 i & \phantom{+}0.5\phantom{i}
\end{bmatrix} \nonumber\\
R_y(\varphi) &=&
\begin{bmatrix}
\phantom{+}0.9 & \phantom{+}0.5 & \phantom{+}0.2 \\
-0.5 & \phantom{+}0.7 & \phantom{+}0.5 \\
\phantom{+}0.9 & -0.5 & \phantom{+}0.9
\end{bmatrix} \\
R_z(\omega) &=&
\begin{bmatrix}
-0.5 - 0.9\,i & 0 & 0 \\
0 & 1 & 0 \\
0 & 0 & -0.5 + 0.9\,i
\end{bmatrix} .\nonumber
\label{eq:robst}
\end{eqnarray}
By applying the three matrices to the state $\ket{\psi}$ calculated in Sec.~\ref{MajorabaRobust}, the following state is obtained.
\begin{equation}
R_z(\omega)R_y(\varphi)R_x(\varepsilon)\ket{\psi} = \ket{\phi}=\begin{bmatrix}
-0.34-0.66i \\
-0.54+0.29i \\
\phantom{+}0.06+0.25i
\end{bmatrix} .
\label{eq:MajorabaRobust}
\end{equation}
Implementing it in the circuit, the result matches the mathematical expression from Eq.~(\ref{eq:MajorabaRobust}), thus verifying the realisation of these new rotation gates applied to Majorana points. Further tests combining Majorana encoding, different rotations, and circuits with more than one qutrit were also carried out, for all of which the circuit output matches the mathematical expression.
\begin{figure*}[!t]
  \centering
  \subfloat[\justifying Normalized distributions of the $N$-subjettiness variable $\tau_1$ for background (QCD) and signal jets ($H\to b\bar{b}$, $W\to q\bar{q}$, and $t\to bq\bar{q}$). ]{%
    \includegraphics[width=0.31\textwidth]{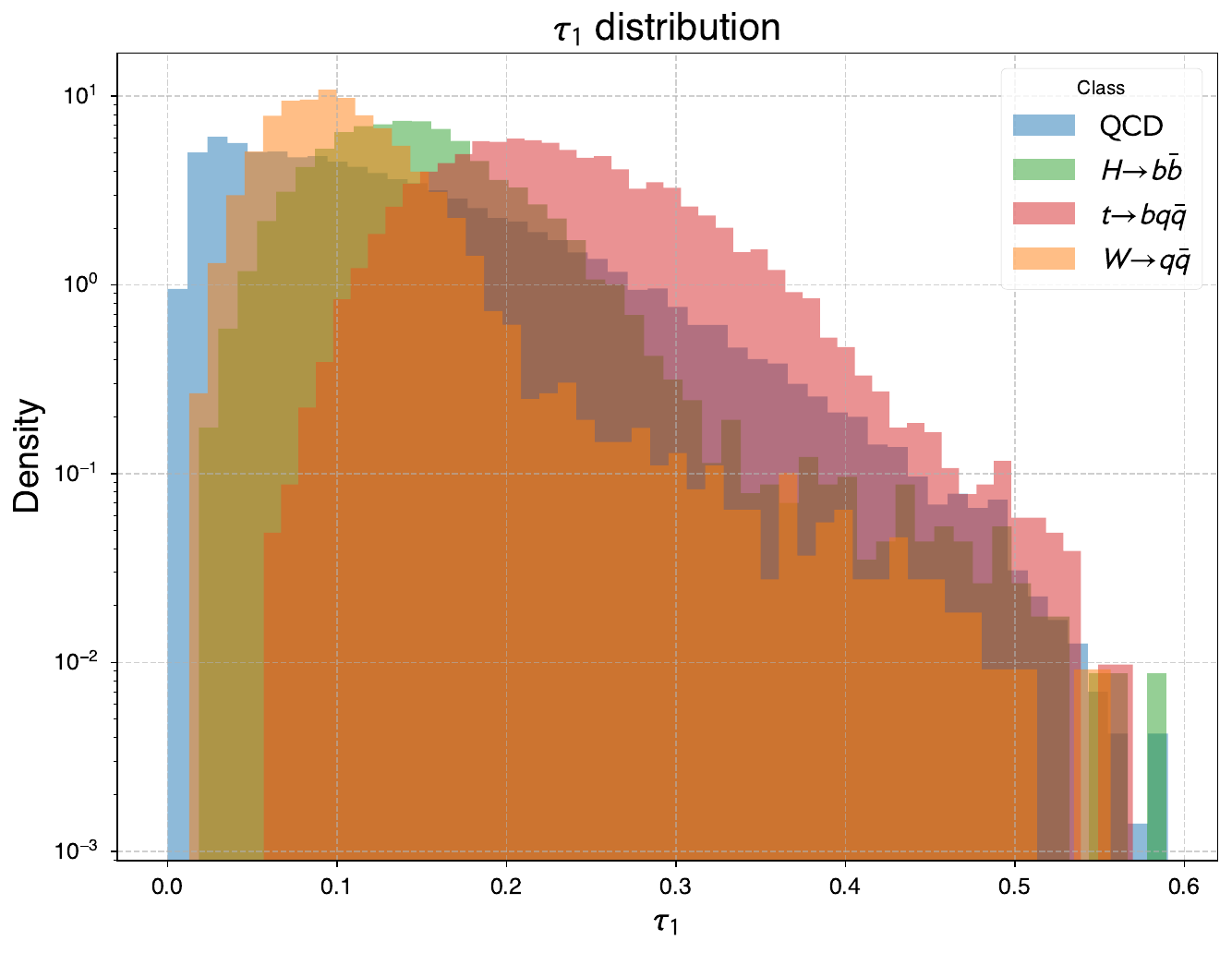}
    \label{a}
  }
  \hfill
  \subfloat[\justifying Normalized distributions of the $N$-subjettiness variable $\tau_2$ for background (QCD) and signal jets ($H\to b\bar{b}$, $W\to q\bar{q}$, and $t\to bq\bar{q}$).]{%
    \includegraphics[width=0.31\textwidth]{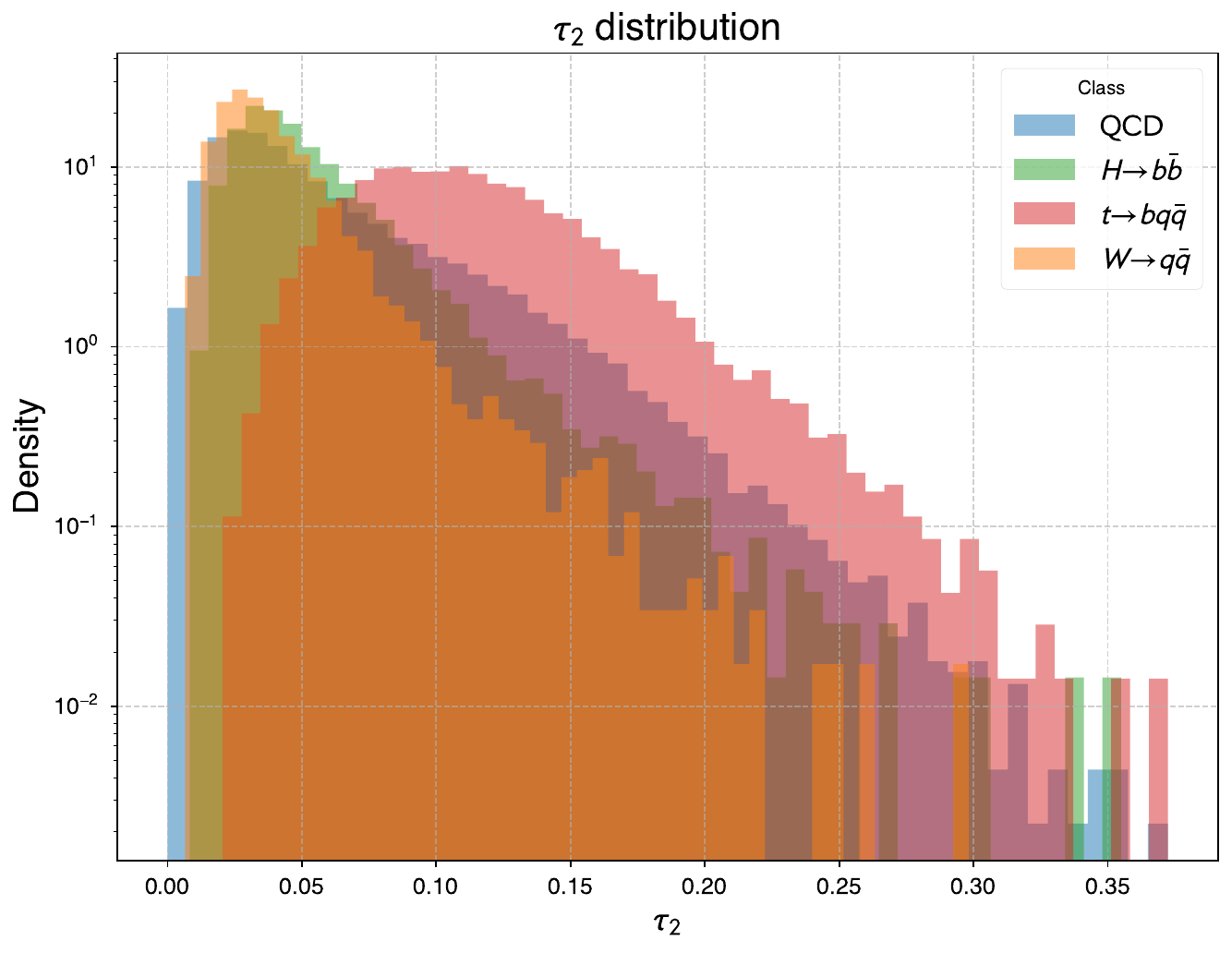}
    \label{b}
  }
  \hfill
  \subfloat[\justifying Normalized distributions of the $N$-subjettiness variable $\tau_3$ for background (QCD) and signal jets ($H\to b\bar{b}$, $W\to q\bar{q}$, and $t\to bq\bar{q}$).]{%
    \includegraphics[width=0.31\textwidth]{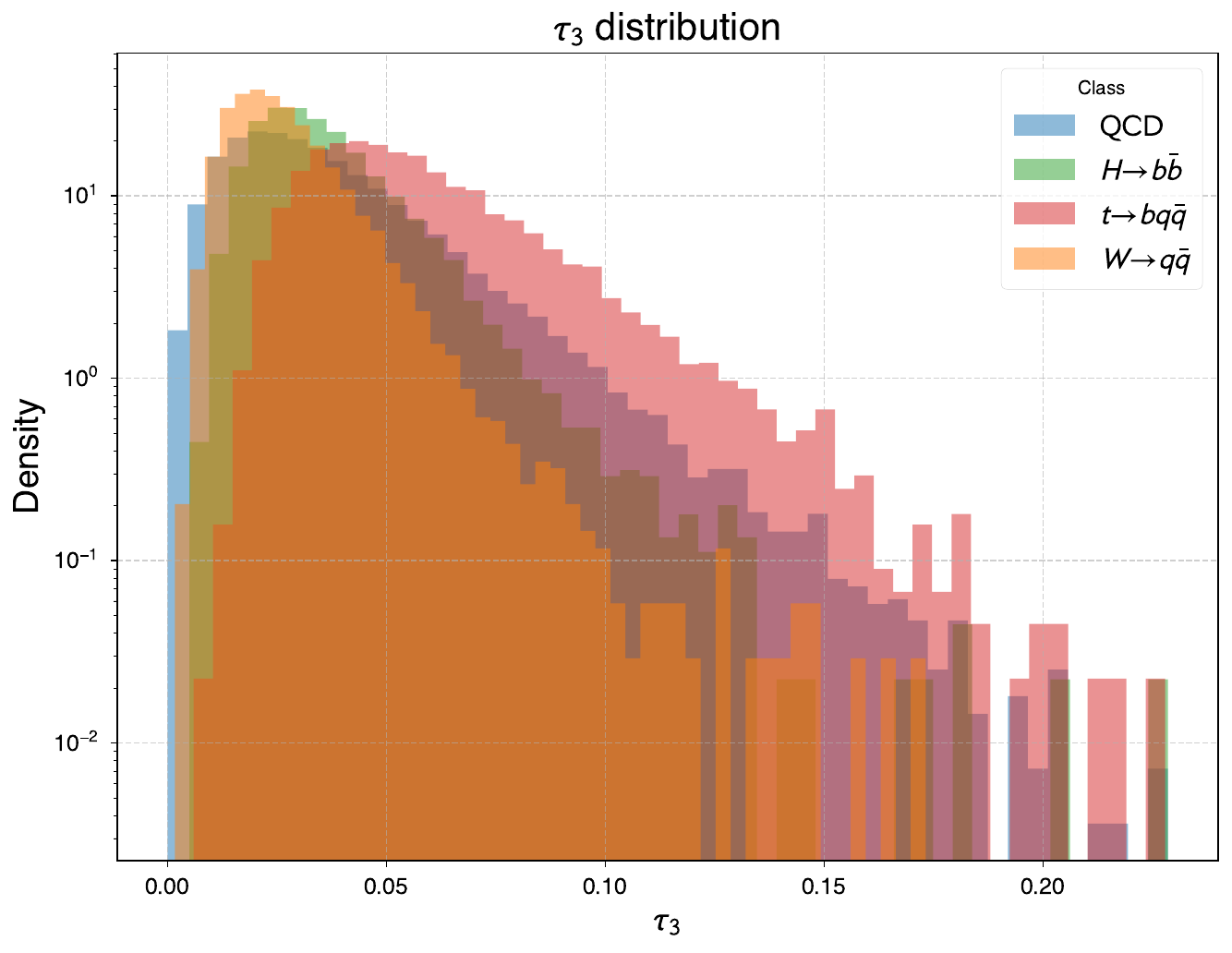}
    \label{c}
  }\\
  \subfloat[\justifying Normalized distributions of the $N$-subjettiness variable $\tau_4$ for background (QCD) and signal jets ($H\to b\bar{b}$, $W\to q\bar{q}$, and $t\to bq\bar{q}$).]{%
    \includegraphics[width=0.31\textwidth]{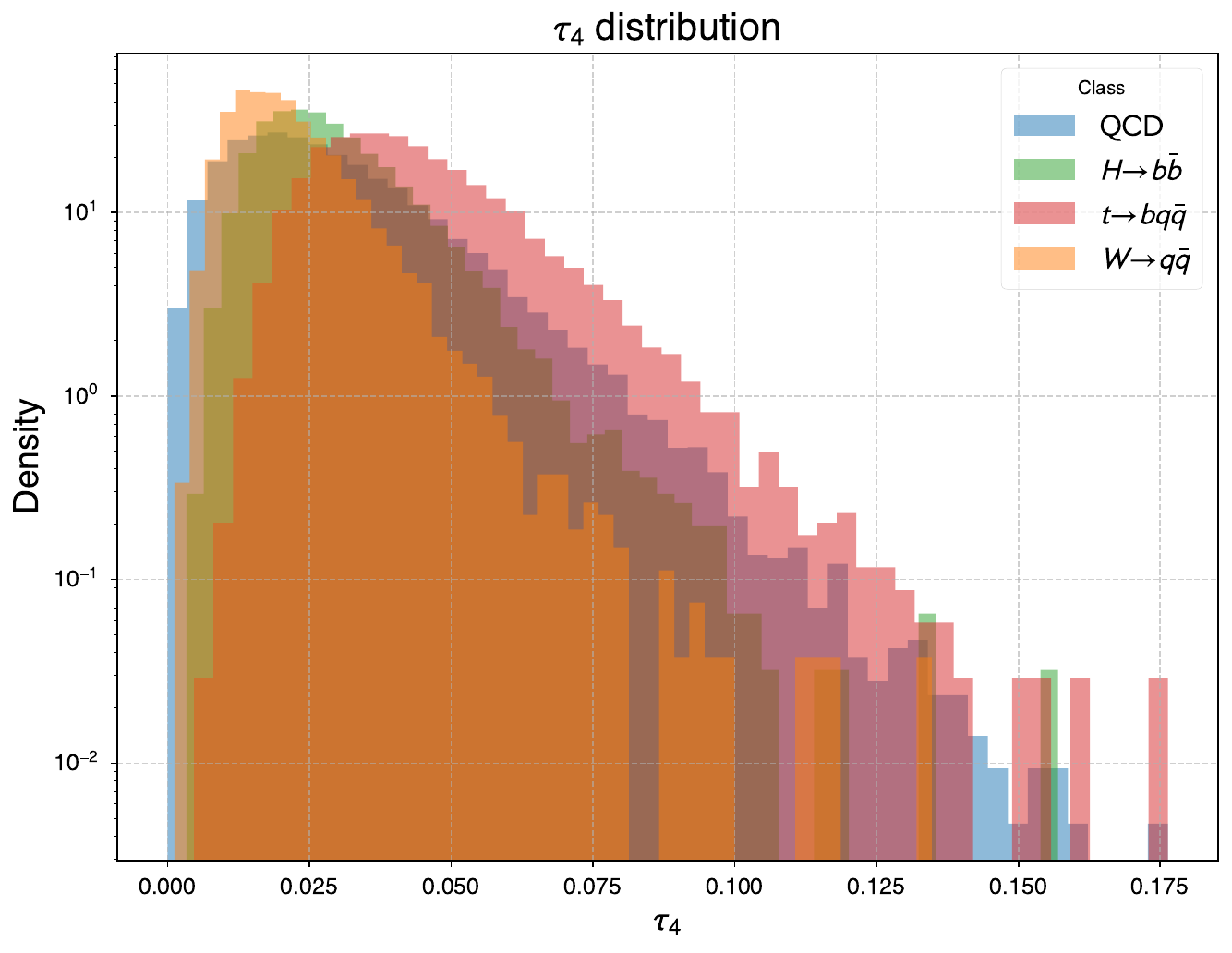}
    \label{d}
  }
  \hfill
  \subfloat[\justifying Normalized distributions of the $\tau$-ratio $\tau_{12}$ for background (QCD) and signal jets ($H\to b\bar{b}$, $W\to q\bar{q}$, and $t\to bq\bar{q}$).]{%
    \includegraphics[width=0.31\textwidth]{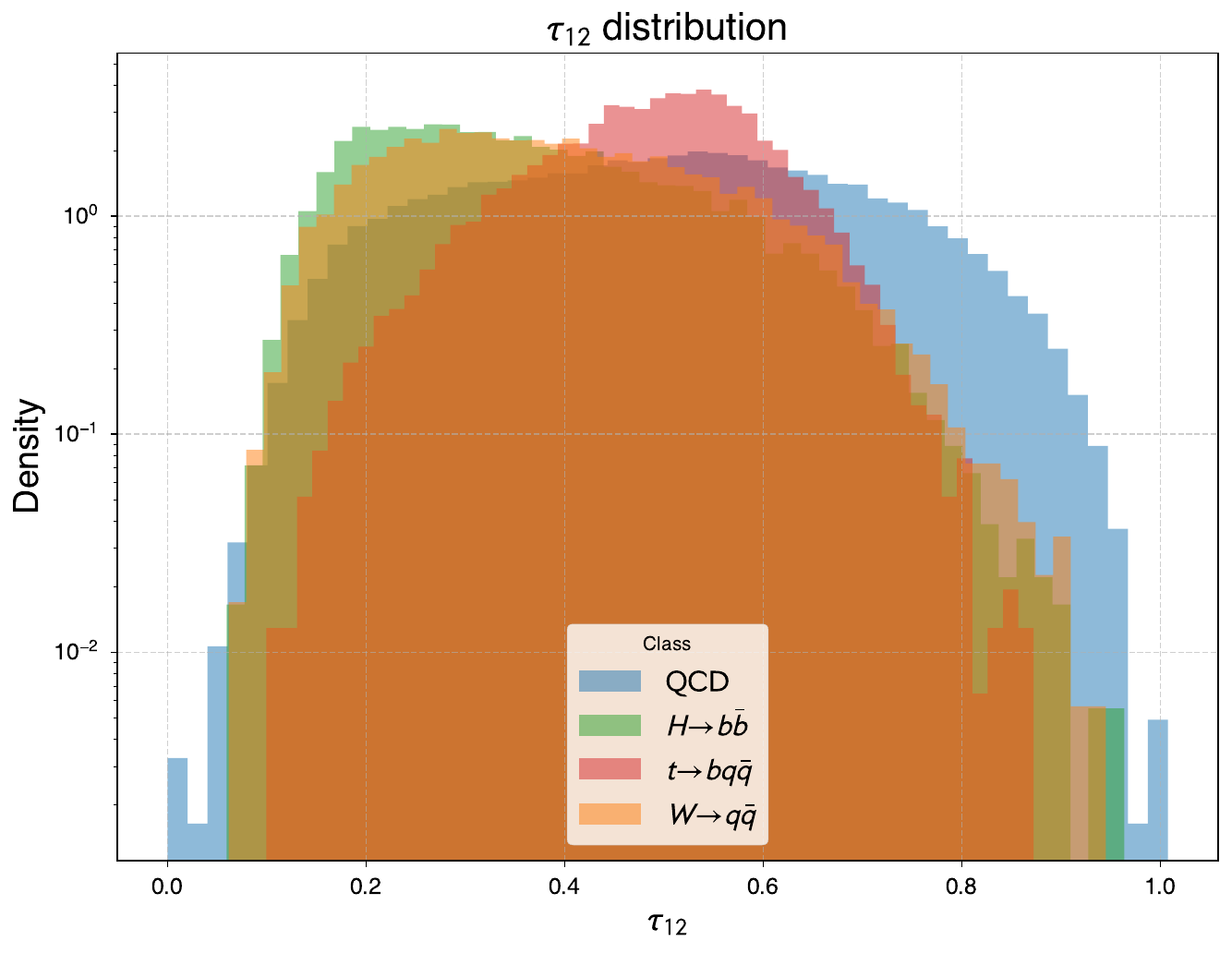}
    \label{e}
  }
  \hfill
  \subfloat[\justifying Normalized distributions of the $\tau$-ratio $\tau_{23}$ for background (QCD) and signal jets ($H\to b\bar{b}$, $W\to q\bar{q}$, and $t\to bq\bar{q}$).]{%
    \includegraphics[width=0.31\textwidth]{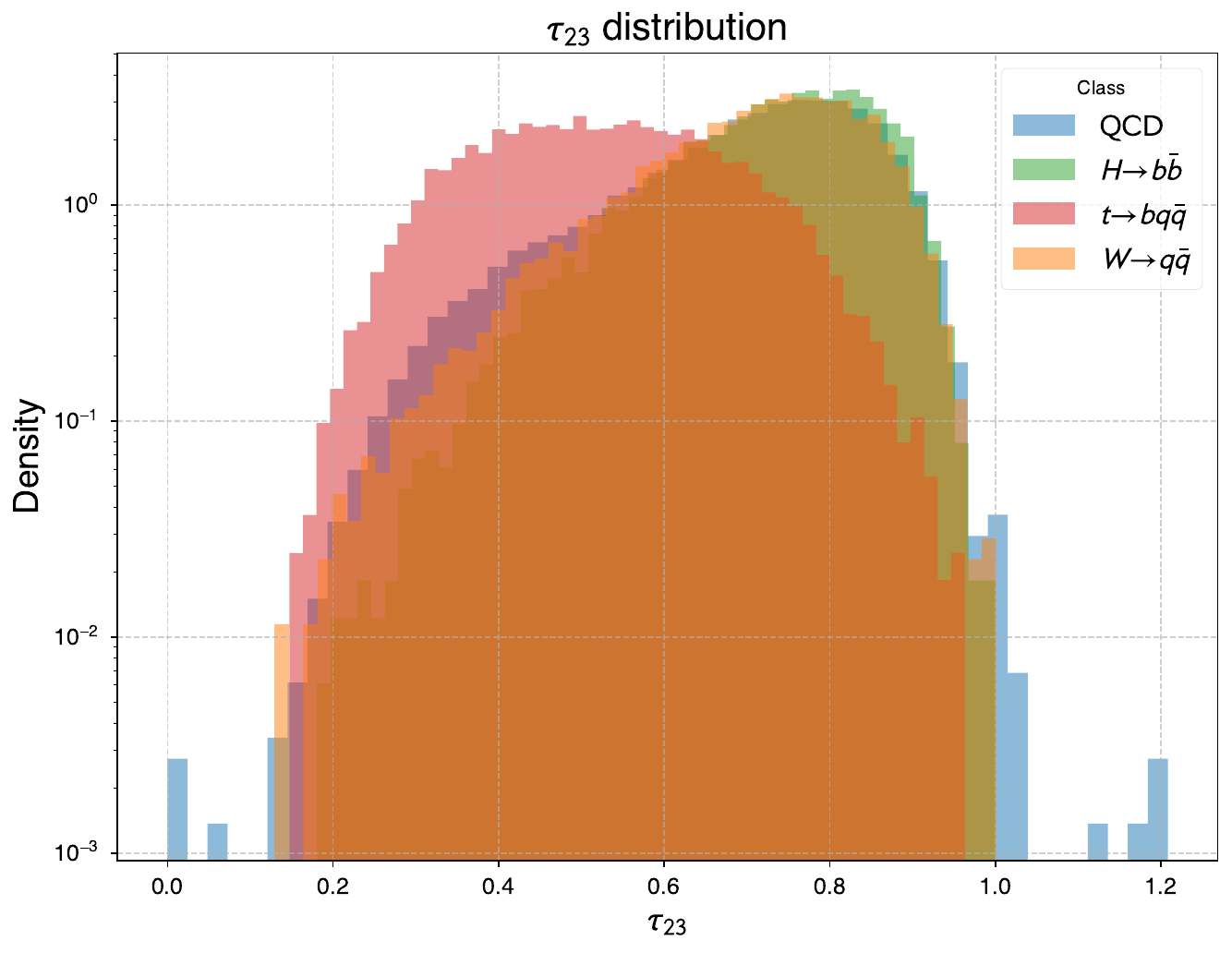}
    \label{f}
  }

  \caption{\justifying \footnotesize \textbf{Normalized $\tau_N$ distributions through the 4 classes} (`QCD', $W \rightarrow q\bar{q}$, $H \rightarrow b\bar{b}$, $t \rightarrow bq\bar{q}$). The distributions are shown on a logarithmic scale to highlight the differences in the tails. (a) $N$-subjettiness variable: $\tau_1$, (b) $N$-subjettiness variable: $\tau_2$, (c) $N$-subjettiness variable: $\tau_3$, (d) $N$-subjettiness variable: $\tau_4$, (e) $\tau$-ratio: $\tau_{12}$, (f) $\tau$-ratio: $\tau_{23}$.}
  \label{fig:taus}
\end{figure*}
\section{\label{app:dis} Jensen-Shannon distance}
To evaluate the similarity between the different fidelity distributions, the square root of the Jensen-Shannon (JS) divergence ($\sqrt{\textrm{JS}}$)  is used as a metric, as it represents a more stable and accurate method to calculate the divergence between two distributions~\cite{jensen,jensen2} when compared to other standard metrics such as the Kullback–Leibler divergence. 

The JS divergence measures the deviation between the Shannon entropy ($H$) of the mixture $(P+R)/2$ and the mixture of the entropies, and it is defined for all probability distributions. It is bounded and symmetric, and vanishes only when $P=R$~\cite{jensen}.
Its definition is given by
\begin{eqnarray}
\mathrm{JS}(P,R) = H\!\left(\frac{P + R}{2}\right) - \frac{1}{2}\bigl(H(P) + H(R)\bigr).\nonumber\\
\end{eqnarray}
The $\sqrt{\textrm{JS}}$ distance thus defined is closer to zero the more similar the distributions are~\cite{jensen,jensen2}. In the context of this work, the $\sqrt{\textrm{JS}}$ distance is used as a metric to evaluate the expressive power of our model by measuring the distance between the different fidelity distributions. Larger values of $\sqrt{\textrm{JS}}$ indicate a greater capacity to discriminate between distinct physical signals, as well as a higher level of expressive power.

\section{\label{app:wass} Wasserstein distance} 
To evaluate the model’s ability to discriminate between distinct signals and to quantify the similarity between fidelity distributions, we also employ the $W_1$ Wasserstein distance~\cite{wassertein_metric}. Also referred to as the Earth Mover’s Distance (EMD), this metric offers a rigorous framework for comparing probability distributions (see Ref.~\cite{wassertein_metric} for more details). The Wasserstein distance between two probability distributions $p_1$ and $p_2$ is defined as:
\begin{equation}
    W(p_1, p_2) = \inf_{\gamma \in \Pi(p_1, p_2)} \mathbb{E}_{(x, x') \sim \gamma} \left[ \|x - x'\| \right],
    \label{eq:wasserstein}
\end{equation}
where $\inf$ denotes the infimum (the greatest lower bound) of the set, $\Pi(p_1, p_2)$ represents the set of all joint distributions (or transport plans) $\gamma$ whose marginals are $p_1$ and $p_2$, and $\mathbb{E}_{(x, x') \sim \gamma}$ is the expected value of the distance between samples $x$ and $x'$ under the coupling $\gamma$.
Intuitively, the Wasserstein distance represents the cost of the optimal transport plan needed to move one distribution into another. It is particularly effective when the distributions have markedly different shapes~\cite{Leo2023Wasserstein}, since the cost is computed as the product of the probability mass that is moved and the distance over which it is transported.

Consequently, this geometric awareness has led to its successful application in various machine learning domains, including generative models and domain adaptation, where it frequently outperforms Euclidean distance and Kullback–Leibler divergence~\cite{wessertein_2,wessertein_2, cms_wassertein,arjovsky2017principledmethodstraininggenerative,gulrajani2017improvedtrainingwassersteingans,rosca2017variationalapproachesautoencodinggenerative, NIPS2015_a9eb8122, Blanchet_Kang_Murthy_2019}. 

Beyond simple magnitude differences, the Wasserstein distance captures the underlying structure and geometry of the data distributions. In contrast to traditional metrics that only consider pointwise differences, this approach incorporates the location and correlation between data points.  

In this work, the Wasserstein distance is used as a metric to evaluate the expressive power of our model, measuring the distance between fidelity distributions. Larger Wasserstein distances indicate a greater capacity to discriminate between distinct physical signals. The use of the Wasserstein distance as a loss function is also a possible way to train the model and is left for future work.

\section{\label{app:taus} \texorpdfstring{$\tau_N$ }{tauN} distribution}
The $\tau_N$ parameters quantify the degree to which a jet can be regarded as being composed of $N$ subjets. In this way, jets with $\tau_N \approx 0$ have all their radiation aligned with the candidate subjet directions and therefore have $N$ or fewer subjets. Jets with $\tau_N \gg 0$ have a large fraction of their energy distributed away from the candidate subjet directions and thus contain at least $N+1$ subjets~\cite{taus}.

As illustrated in Supplementary Fig.~\ref{b}, both W  and QCD jets tend to have a small $\tau_2$. Similarly, even though W jets are more likely than QCD jets to have large $\tau_1$, QCD jets with a diffuse spray of large-angle radiation can also show large $\tau_1$. However, those QCD jets with large $\tau_1$ typically also have large values of $\tau_2$, and that is why the ratio $\tau_2/\tau_1$ is preferred as a discriminating variable (see Ref.~\cite{taus} for more details). As shown in Supplementary Fig.~\ref{e}, H and W jets have smaller $\tau_2/\tau_1$ values than QCD jets and, in the same way, $t$ jets have smaller values of $\tau_3$ but higher values of $\tau_2$.  Supplementary Figure~\ref{f} shows the clear differentiation in the $t$-class distribution.

\section{\label{app:wass} $p_T$ distributions} 

\begin{figure*}[!t]
  \centering
  \subfloat[\justifying The $p_T$ distributions of all constituents versus those of the four leading constituents for background (q/g) events in the JetClass simulated CMS dataset.]{
    \includegraphics[width=0.45\textwidth]{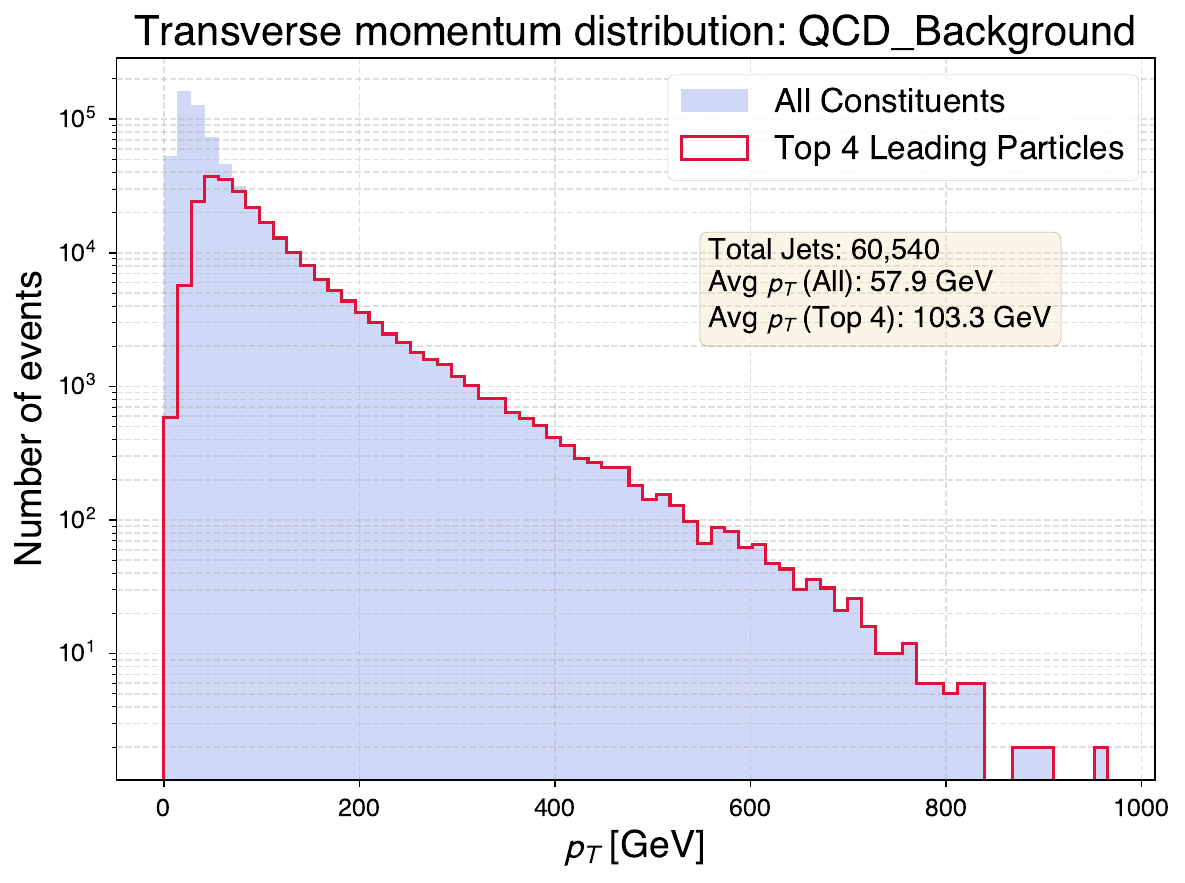}
    \label{fig:qubits_R}
  }
  \hfill
  \subfloat[\justifying The $p_T$ distributions of all constituents versus those of the four leading constituents for $H \to b\bar{b}$ events in the JetClass simulated CMS dataset.]{
    \includegraphics[width=0.45\textwidth]{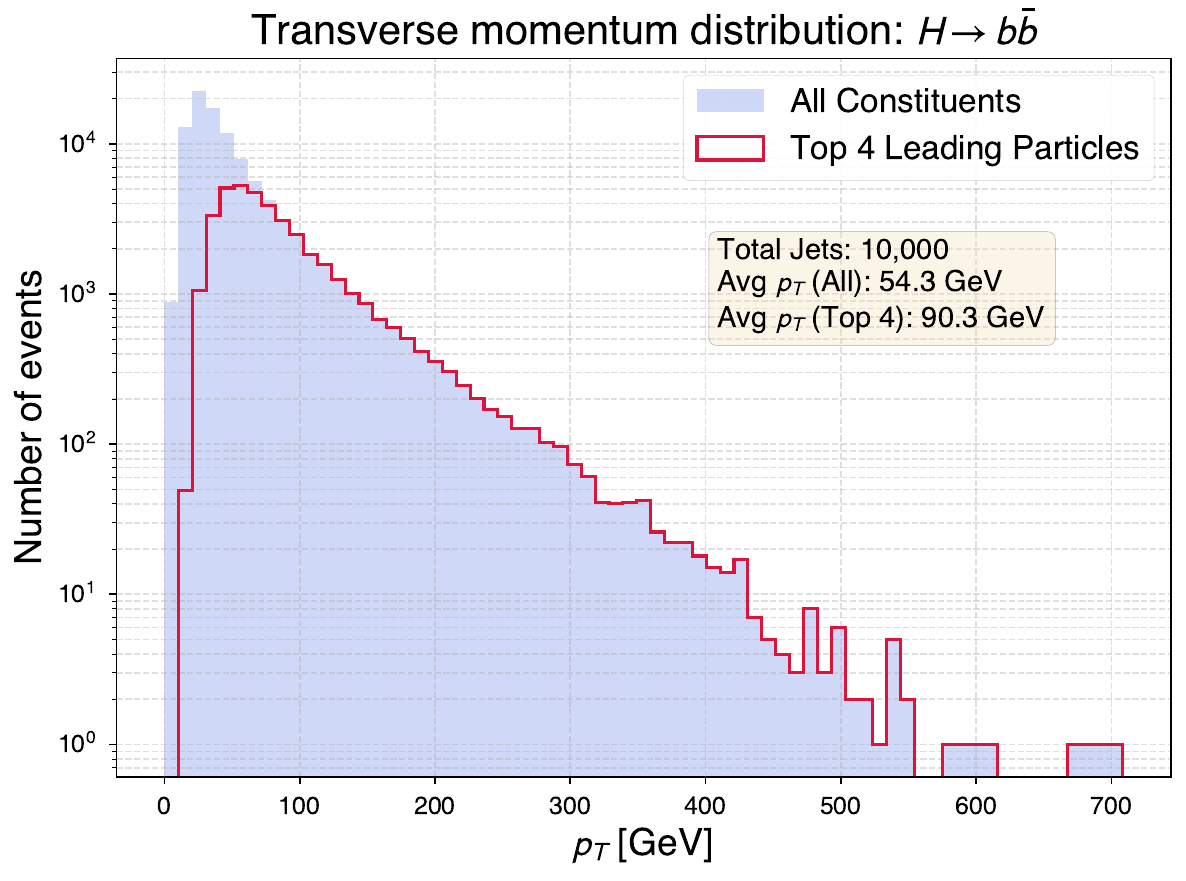}
    \label{fig:qutrits_R}
  }

  \medskip 

  \subfloat[\justifying The $p_T$ distributions of all constituents versus those of the four leading constituents for $W \to q\bar{q}$ events in the JetClass simulated CMS dataset.]{
    \includegraphics[width=0.45\textwidth]{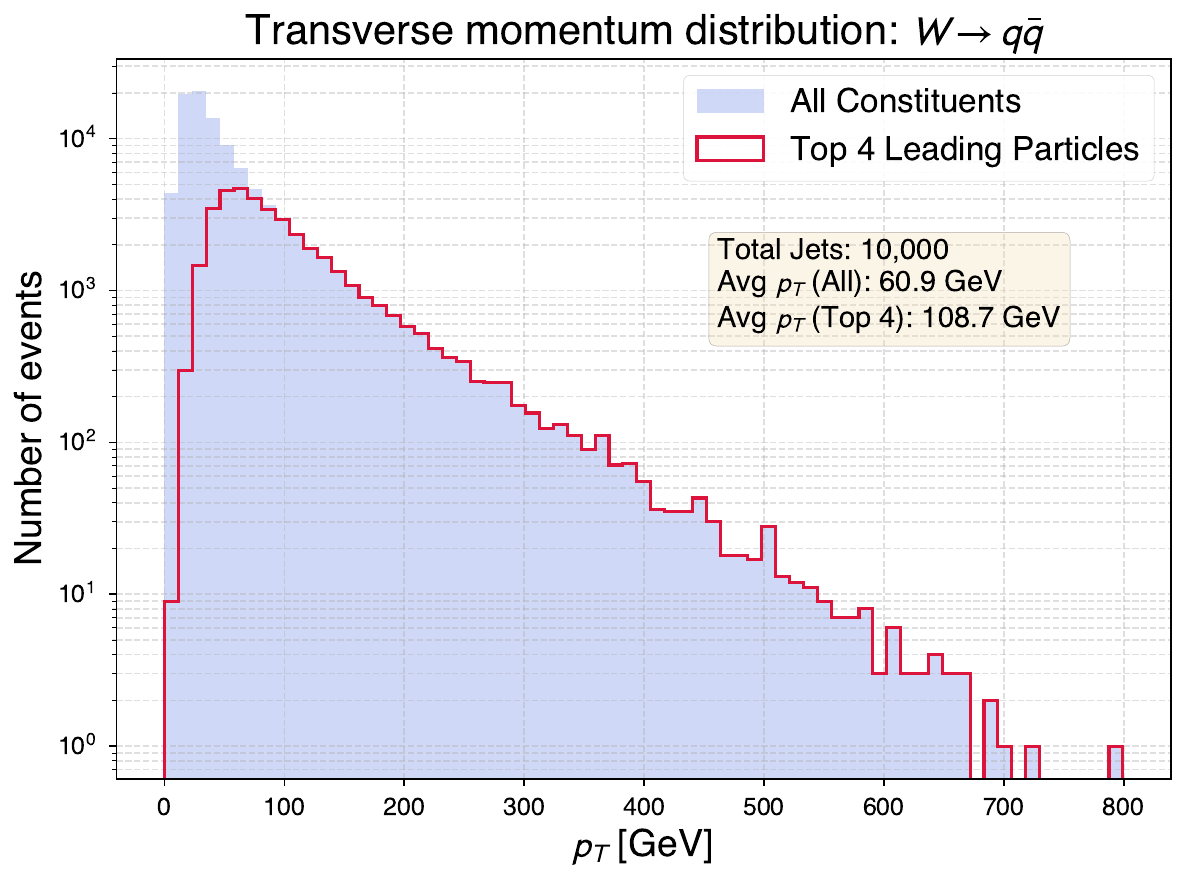}
    \label{fig:12E}
  }
  \hfill
  \subfloat[\justifying The $p_T$ distributions of all constituents versus those of the four leading constituents for $t \to bq\bar{q}$ events in the JetClass simulated CMS dataset.]{
    \includegraphics[width=0.45\textwidth]{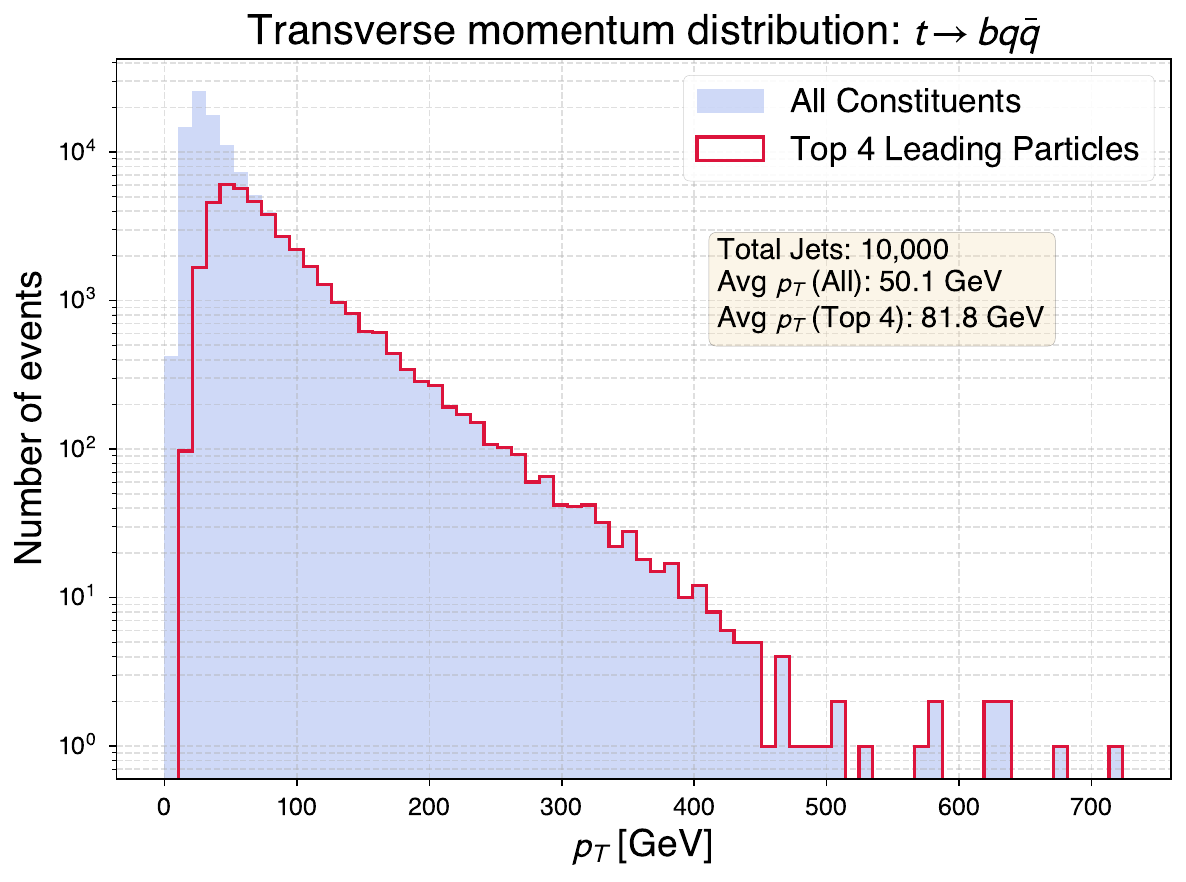}
    \label{fig:qubits_s}
  }
\caption{\protect\footnotesize\label{fig:wide}\protect\justifying \textbf{Histogram comparing the $p_T$ distributions of all constituents in the simulated CMS dataset with the $p_T$ distributions of the four leading constituents.} The plot also includes the total number of jets, as well as the average values for both cases: all constituents and the four leading particles.}
\end{figure*}

Histograms comparing the $p_T$ distributions of all constituents in the simulated CMS dataset with the $p_T$ distributions of the four leading constituents are plotted in Supplementary Fig.~\ref{fig:wide} as a way to analyse the restriction to only four particles per jet due to memory limitations. As can be seen, the leading constituents capture the dominant energy flow and the characteristic prong-like topology of the signals, which are essential for effective discrimination. While the omission of additional constituents may lead to a loss of soft, wide-angle radiation details, the high-fidelity reconstruction of these four leading particles appears sufficient for the QAE to distinguish between the background norm and the anomalous signal structures.
\clearpage
\newpage
\bibliography{apssamp1}

\end{document}